\documentclass[traditabstract,a4]{aa}  

\usepackage{epsfig}
\usepackage{graphicx}
\usepackage{epstopdf}
\usepackage{color}  
\usepackage{array}   
\usepackage{units}  
\usepackage{natbib}  
\usepackage{multirow}   
\usepackage{amsmath,amssymb}  
\usepackage[english]{babel}
\usepackage[latin1]{inputenc}
\usepackage[T1]{fontenc}
\usepackage{longtable,lscape}
\usepackage{footnote}
\usepackage{txfonts}
\usepackage[toc,page]{appendix} 

\begin{document}

\newcommand{\avg}[1]{\left< #1 \right>}


   \title{Measurement of the $T_{\rm CMB}$ evolution from \\the Sunyaev-Zel'dovich effect} 

   \author{G. Hurier\inst{1}
   \and N. Aghanim\inst{1}
   \and M. Douspis\inst{1}
   \and E. Pointecouteau\inst{2}
          }

\institute{Institut d'Astrophysique Spatiale, CNRS (UMR8617) Universit\'{e} Paris-Sud 11, B\^{a}timent 121, Orsay, France
\and CNRS; IRAP; 9 Av. colonel Roche, BP 44346, F-31028 Toulouse cedex 4, France 
\\Université de Toulouse; UPS-OMP; IRAP; Toulouse, France\\
\\
\email{ghurier@ias.u-psud.fr} 
}

   \date{Received /Accepted}
 
   \abstract{ In the standard hot cosmological model, the black-body
     temperature of the cosmic microwave background (CMB), $T_{\rm
       CMB}$, increases linearly with redshift.  Across the line of
     sight CMB photons interact with the hot ($\sim10^{7-8}$~K ) and
     diffuse gas of electrons from galaxy clusters.  This interaction
     leads to the well-known thermal Sunyaev-Zel'dovich effect (tSZ),
     which produces a distortion of the black-body emission law,
     depending on $T_{\rm CMB}$.  Using tSZ data from the {\it Planck}
     satellite, it is possible to constrain $T_{\rm CMB}$ below z=1.
     Focusing on the redshift dependance of $T_{\rm CMB}$, we obtain
     $T_{\rm CMB}(z)=(2.726\pm0.001)\times (1+z)^{1-\beta}$~K with
     $\beta=0.009\pm0.017$, which improves on previous constraints. Combined
     with measurements of molecular species absorptions, we derive
     $\beta=0.006\pm0.013$. These constraints are consistent with the
     standard (i.e. adiabatic, $\beta=0$) Big-Bang model.}

   \keywords{Cosmology: observations -- Cosmic background radiation --
     Galaxies: clusters: general}

   \maketitle


\section{Introduction}

The cosmic microwave background (CMB) radiation is a
fundamental observational probe of the hot Big-Bang model. In standard $\Lambda {\rm CDM}$ 
cosmology, the CMB black-body temperature evolution as
a function of redshift reads $T_{\rm CMB}(z)~=~T_{\rm
  CMB}(z=0)~(1+z)^{1-\beta}$, with $\beta=0$ and $T_{\rm
  CMB}(z=0)$ is the temperature measured today in the local Universe.\\

A violation of this evolution would imply deeper theoretical
consequences \citep[e.g.,][]{uza04}.  Alternative cosmological models
predict a nonlinear scaling law between temperature and redshift
\citep[see e.g.,][]{maty95,ove98,puy04}.  A deviation from $\beta=0$ can be induced by
 a violation of the equivalence
principle or by a non-conservation of the photon number without any spectral
distortion of the CMB.  In the first case, it can be associated
with fundamental constant variations \citep[see
  e.g.,][]{mur03,sri04}.  In the second case, it can be associated with
decaying dark-energy models \citep[see e.g.,][]{fre87,lim96,jer11} or
with axion-photon-like coupling process \citep[see][for a recent
  review]{jae10}.\\

The CMB temperature has been measured in our galaxy at high precision
using CN excitation in molecular clouds by \citet{rot93}. Next, it has
been measured using the COBE-FIRAS experiment \citep{fix96,mat99} and
more recently, using the WMAP data \citep{ben03} to recalibrate
the COBE-FIRAS \citep{fix09} measurements.\\ 
In addition to the
measurement of $T_{\rm CMB}$ at $z=0$, there are currently two direct
observational methods to measure $T_{\rm CMB}$ at redshifts $z>0$. The
first one uses the excitation of interstellar atomic or molecular
species by CMB photons \citep[e.g.][]{los01}. This approach was
used even before the CMB discovery \citep[see][for a review of early
  observations]{tha72}.  When in radiative equilibrium with the CMB radiation, the
excitation temperature of molecular species equals that of the CMB.  
Thus, these species allow one to measure the CMB temperature
\citep[][]{bah68}.  The fact that the
CMB is not the only heating source of the interstellar medium and the
lack of a detailed knowledge of the physical conditions in the
absorbing clouds are the main sources of systematics and
uncertainties \citep{com99}. They lead to upper limits or large error
bars ($\Delta T \ge 0.6$~K) on the value of $T_{\rm CMB}$
\citep{mey86,son94b,son94a,lu96,rot99}. Recently, \citet{mul13} have 
performed a comprehensive analysis that overcomes parts of the limitations.
In particular, they have made use of various atomic and molecular species
  simultaneously in order to completely model the observed 
cloud properties and hence to provide tight constraints on $T_{\rm CMB}$.\\

The second observational approach consists in measuring a weak
spectral distortion of the CMB black-body in the direction of galaxy
clusters that is caused by the thermal Sunyaev-Zel'dovich (tSZ) effect
\citep{zel69,sun72}. This technique was originally proposed by
\citet{fab78} and \citet{rap80}. Predictions for these
measurement with {\it Planck} have been discussed by \citet{hor05} and
\citet{mar12}. They predicted, in an optimistic case, an uncertainty 
on the $T_{\rm CMB}$ evolution of $\Delta \beta \sim 0.011$. Actual measurements
were already performed \citep{bat02,luz09} and provided
$\Delta T \leq 0.3$~K on a small number of clusters of galaxies. The {\it
  Planck} satellite now offers a new and large sample of galaxy
clusters \citep{PlanckSZC} observed via the tSZ effect that allows us
to derive much tighter constraints on the evolution of the CMB temperature. 

Galaxy cluster catalogs that reach deeper (e.g. SPT, \citet{rei13} and ACT,
\citet{has13}) or larger (e.g. MCXC, \citet{pif11}) can also be used to complement the constraint from the {\it
  Planck} SZ clusters.

In this work, we focus on improving in the measurement of
$T_{\rm CMB}$ at redshifts $z>0$ from the {\it Planck} data using the
tSZ effect from galaxy clusters.  In Sect.~\ref{sec_data} we present
the {\it Planck} intensity maps and galaxy cluster catalog that were
used. Sect.~\ref{sec:theory} briefly presents the tSZ effect. Then in
Sect.~\ref{sec_meth}, we present the stacking method used to extract
the tSZ flux in the different {\it Planck} frequency channels to
constrain $T_{\rm CMB}$. In Sect.~\ref{sec_error}, we estimate the
uncertainty levels on our measurement that are caused, on the one hand, by
foreground/background contributions to {\it Planck} intensity maps
and, on the other hand, by instrumental systematic effects. Next in
Sect.~\ref{sec_ana}, we carefully model our measurement to determine
$T_{\rm CMB}$ from the tSZ spectral law. In Sect.~\ref{sec_res}, we
present our results and compare them with previous
measurements. Finally in Sect.~\ref{sec_disc}, we discuss our results
and their cosmological implications.

\section{Data}
\label{sec_data}

\subsection{ {\it Planck} intensity maps}
\label{planckmaps}

We used the six channel maps, 100 to 857~GHz, from
the {\it Planck} satellite \citep{PlanckMIS}.  We refer to
the \citet{PlanckDPC} and \citet{PlanckCAL} for the generic scheme of time
ordered information processing and map-making, as well as for the
characteristics of the {\it Planck} frequency maps.  We used the {\it
  Planck} channel maps in {\tt HEALPix} \citep{gor05}
$N_{\mathrm{side}}=2048$ at full resolution.  Here, we approximate the
{\it Planck} beams by effective circular Gaussians \citep[see][for
  more details]{PlanckBEAM}.

\subsection{ {\it Planck} SZ catalog}
\label{plcat}
We also used the {\it Planck} catalog of tSZ detections 
\citep[PSZ, ]{PlanckSZC}. 
It contains 1227 sources, which is about six times larger than the 
{\it Planck} Early SZ (ESZ) \citep{PlanckESZ} sample and is currently the largest 
tSZ-selected catalog. The PSZ contains 861 confirmed clusters to 
date.\\
In the following, we only consider the sample of of 813 {\tt
  Planck} cluster with redshifts up to $\sim 1$ 
\citep{PlanckHZ}. This sample 
consists of high-significance SZ detections in the {\it Planck} channel 
maps and thus allows us to constrain $T_{\rm CMB}$ up to $z\sim1$, avoiding
strong contamination by point sources or galactic emission.

\begin{savenotes}
\begin{table}
\caption{Main characteristics of the galaxy cluster catalogs. $N_{\rm cl}$ is the number of objects in the catalog,
  $N^*_{\rm cl}$ is the number of clusters in the present work,
  $\bar{z}$ is the mean redshift, $z_{\rm med}$ is the median
  redshift, $\sigma_z$ is the redshift dispersion of the catalog, and
  $f_{\rm sky}$ is the sky coverage of the catalog.}
\label{tab_cat}
\centering
\begin{tabular}{|c|c|c|c|c|c|c|}
\hline
catalog & $N_{\rm cl}$ & $N^*_{\rm cl}$ & $\bar{z}$ & $z_{\rm med}$ & $\sigma_z$ & $f_{\rm sky}$ \\
\hline
 {\it Planck} SZ & 1227 & 813 & 0.247 & 0.220 & 0.159 & 0.84 \\
MCXC & 1743 & 823 & 0.222 & 0.166 & 0.172 & \phantom{0}---\footnote{The MCXC is a meta-catalog based on ROSAT All Sky Survey catalogs 
\citep[see ][ and reference therein]{pif11}.} \\
SPT & \phantom{0}224 & 142 & 0.574 & 0.576 & 0.279 & 0.02 \\
ACT & \phantom{00}91 & \phantom{0}61 & 0.586 & 0.570 & 0.267 & 0.01 \\
\hline
\end{tabular}
\end{table}
\end{savenotes}

\subsection{Other catalogs of galaxy clusters}
\label{sec_cat}
We complemented our analysis of the {\tt Planck} SZ-cluster 
catalog using X-ray selected clusters from the Meta Catalog 
of X-ray Clusters (MCXC) \citep{pif11}; and tSZ selected clusters from
the Atacama Cosmological Telescope (ACT) \citep{hin10,has13} and the 
South Pole Telescope (SPT) tSZ catalogs \citep{cha09,son12,rei13}. 
Table~\ref{tab_cat} summarizes the main characteristics of these catalogs.
In particular, for the present study, we considered subsamples from the MCXC, ACT, 
and SPT catalogs that consist of clusters not included in the 
PSZ catalog; for the MCXC clusters we also imposed that 
$M_{500}\ge 10^{14}\,M_\odot$.

\subsection{Radio-source catalogs}
Emission from radio point sources is a source of contamination in the 
estimation of the tSZ flux \citep{PlanckSZC} that can lead to 
biases in tSZ Compton-$y$ signal \citep[see e.g.][for an example 
on the Virgo and Perseus galaxy clusters]{hur13}.\\
The NRAO VLA Sky Survey (NVSS)  \citep{con98} is a 1.4 GHz continuum 
survey covering the entire sky north of Dec$~ > -40^\circ$. The 
associated catalog of 
discrete sources contains over 1.8 million radio sources. South of 
Dec$~ < -30^\circ$ and at galactic latitudes $|b| > 10^\circ$,
the Sydney University Molonglo Sky Survey (SUMSS)  \citep{mau03,mau08} 
is a 843 MHz continuum survey also providing a radio source catalog. 
SUMSS and NVSS have a similar sensitivity and angular resolutions, and
combined they cover the whole sky.\\ 
We used these two surveys as tracers of the radio emission 
that contaminated our estimation of the tSZ flux to estimate the 
level of bias in our measure.

\section{The tSZ effect}
\label{sec:theory}

The tSZ effect is a distortion of the CMB black-body radiation 
through inverse Compton scattering. CMB photons receive an average 
energy boost by collision with hot (a few keV) ionized electrons of
 the intra-cluster medium.
The  tSZ Compton parameter in a given direction, $\vec{n}$, on the 
sky is given by
\begin{equation}
y (\vec{n}) = \int n_{e} \frac{k_{\rm{B}} T_{\rm{e}}}{m_{\rm{e}} c^{2} },
\sigma_{\rm T} \  \rm{d}s
\label{comppar}
\end{equation}
where d$s$ is the distance along the line-of-sight,
and $n_{e}$ and $T_{e}$ are the electron number-density and temperature.
In units of CMB temperature, the tSZ effect at a frequency $\nu$ is
\begin{equation}
\frac{\Delta T_{\rm{CMB}}}{T_{\rm{CMB}} }= g(\nu) \ y.
\end{equation}
Neglecting corrections due to the weakly relativistic high-end of 
the velocity distribution for the thermal electrons, we have
\begin{equation}
g(\nu) = \left[ x\coth \left(\frac{x}{2} \right) - 4 \right],
\label{szspec}
\end{equation}
with the dimensionless frequency $ x=h \nu/(k_{\rm{B}} T_{\rm{CMB}})$. $h$ and $k_{\rm B}$ represent the Planck and the Boltzmann constants. 
At $z=0$, where $T_{\rm CMB}(z=0)$~=~2.726$\pm$0.001~K, the tSZ effect
 signal is negative below 217~GHz and positive at higher frequencies.
This characteristic spectral signature is a unique tool 
for detecting of galaxy clusters.
The spectral signature is directly related to $T_{\rm CMB}$ through the
 $x$ variable. They depend on the convolution of the tSZ 
intensity with the {\it Planck} frequency responses.

\begin{figure}[!th]
\center
\begin{tabular}{p{1.2cm}p{1.2cm}p{1.2cm}p{1.2cm}p{1.2cm}}
{100 GHz}&{143 GHz}&{217 GHz}&{353 GHz}&{545 GHz}\\
\multicolumn{5}{c}{\includegraphics[width=8cm]{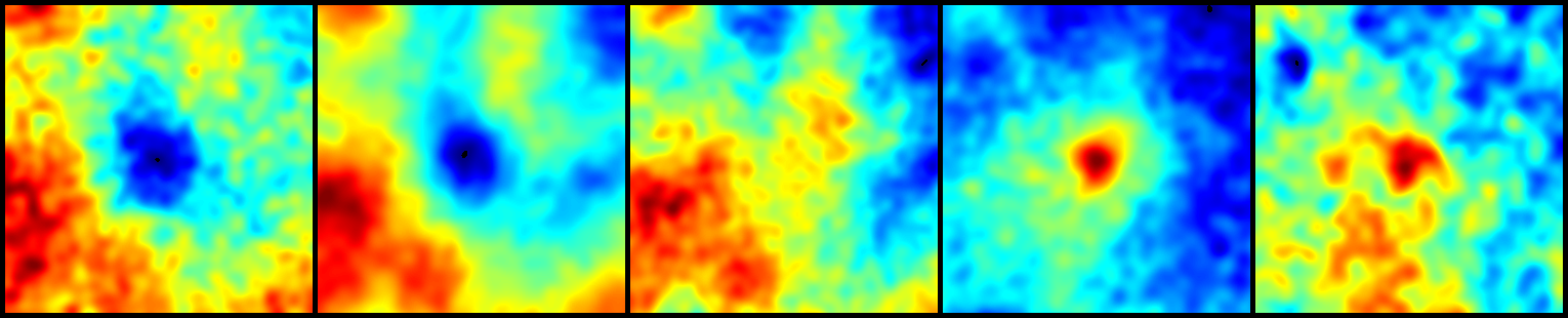}} \\[-0.14cm]
\multicolumn{5}{c}{\includegraphics[width=8cm]{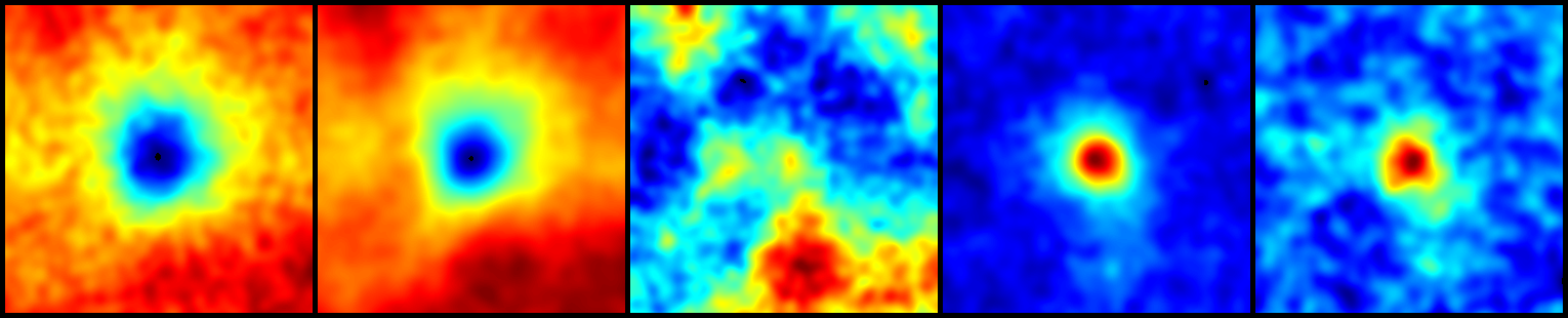}} \\[-0.14cm]
\multicolumn{5}{c}{\includegraphics[width=8cm]{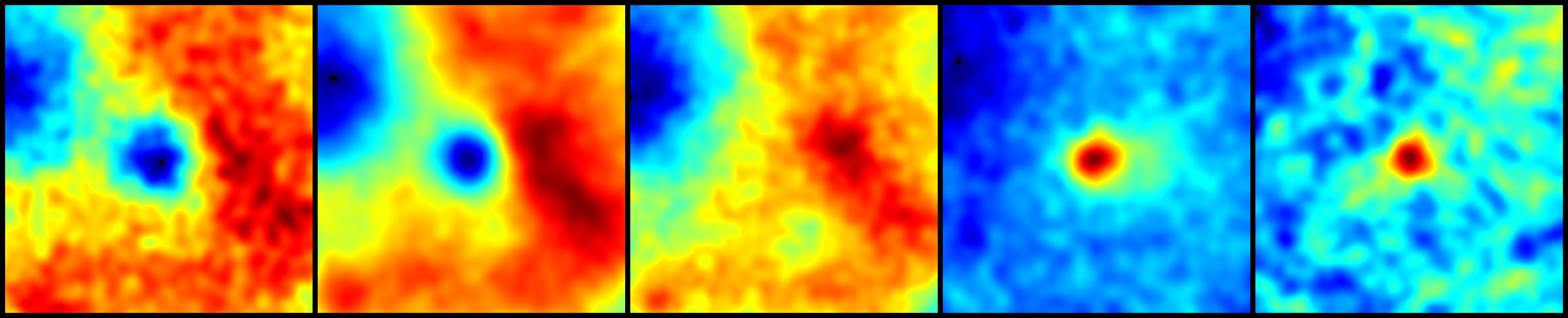}} \\[-0.14cm]
\multicolumn{5}{c}{\includegraphics[width=8cm]{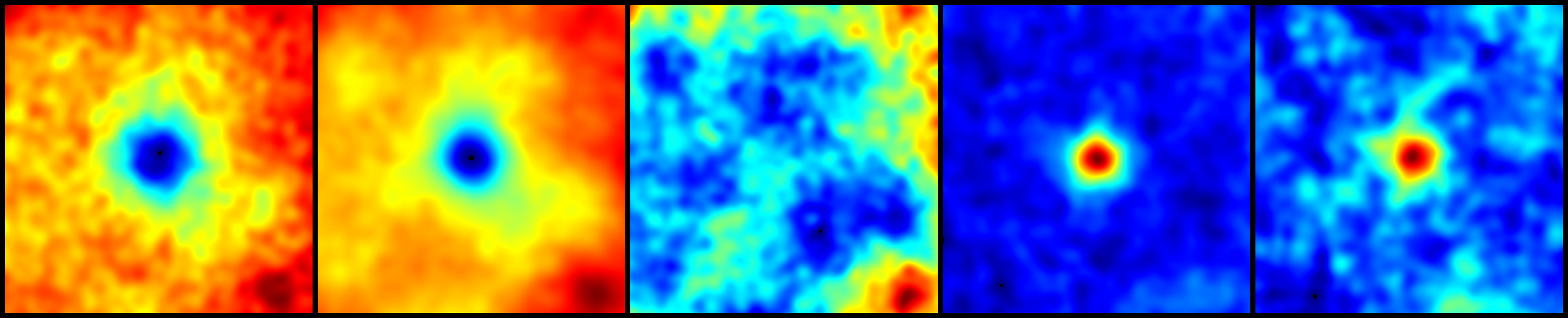}} \\[-0.14cm]
\multicolumn{5}{c}{\includegraphics[width=8cm]{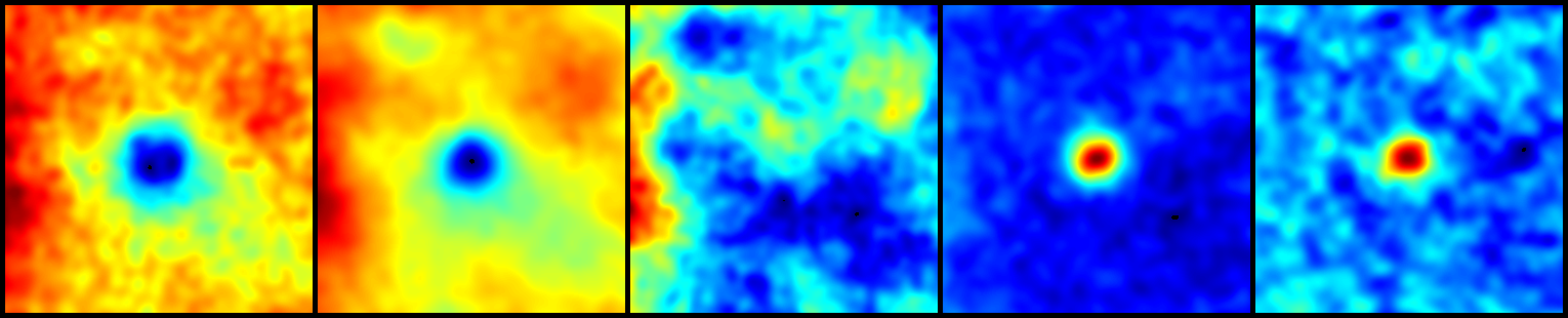}} \\[-0.14cm]
\multicolumn{5}{c}{\includegraphics[width=8cm]{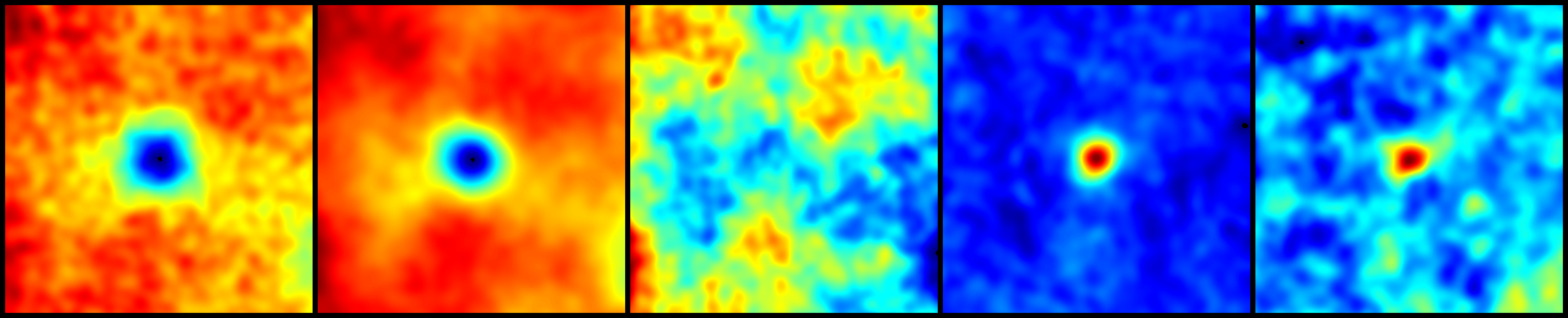}} \\[-0.14cm]
\multicolumn{5}{c}{\includegraphics[width=8cm]{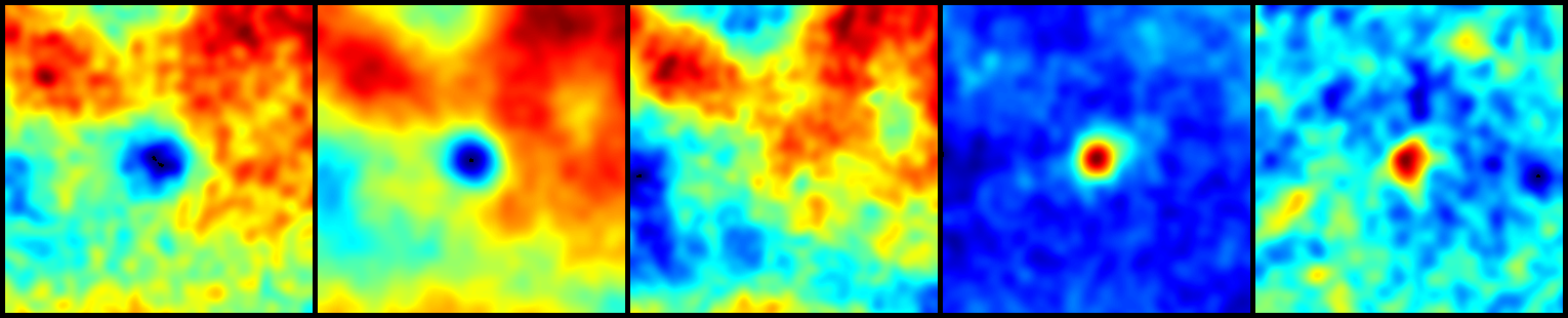}} \\[-0.14cm]
\multicolumn{5}{c}{\includegraphics[width=8cm]{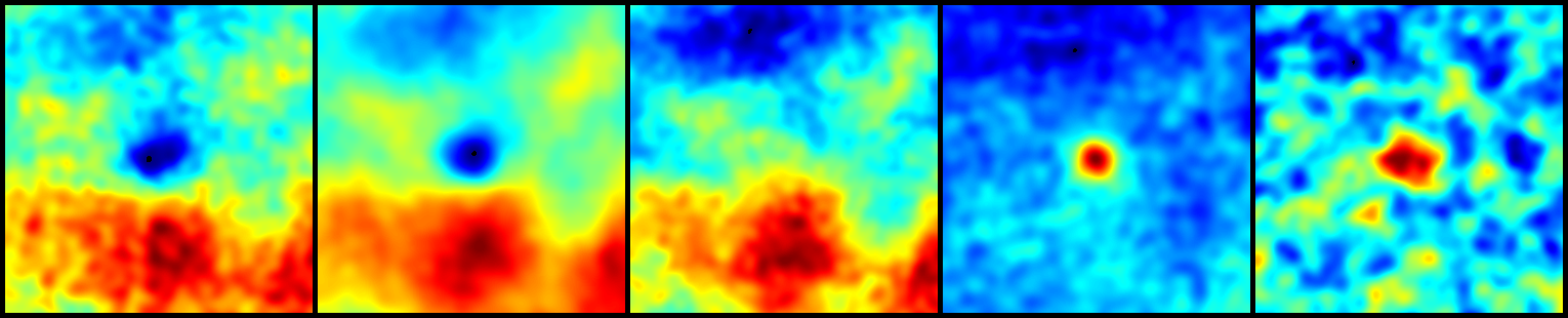}} \\[-0.14cm]
\multicolumn{5}{c}{\includegraphics[width=8cm]{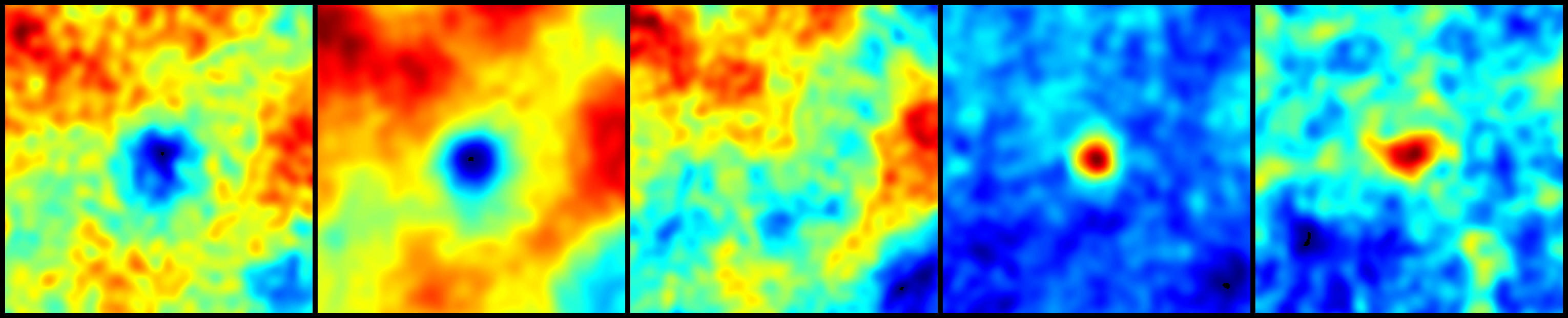}} \\[-0.14cm]
\multicolumn{5}{c}{\includegraphics[width=8cm]{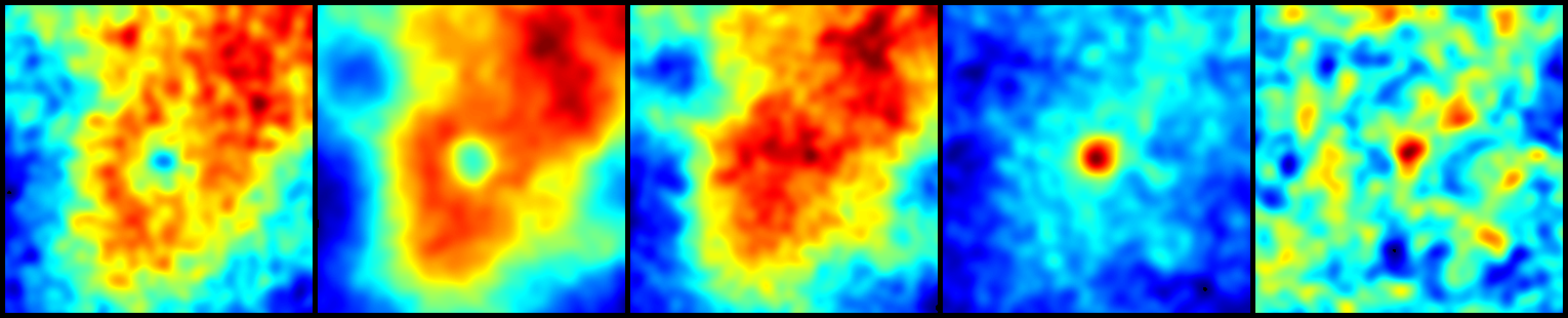}} \\[-0.14cm]
\end{tabular}

\caption{From left to right: stack of {\it Planck} intensity maps from
  100 to 545 GHz cleaned by the 857 GHz channel, centered on the
  location of {\it Planck} tSZ detected clusters. From top to bottom:
  stacked signal from tSZ detected clusters in different redshifts
  bins (from $z=0.0$ to $z=0.5$), see Table~\ref{tabdata} for the
  redshift bins. Each stacked map represents an area
  of 2$^\circ \times$ 2$^\circ$.}
\label{stack1}
\end{figure}

\begin{figure}[!th]
\center
\begin{tabular}{p{1.2cm}p{1.2cm}p{1.2cm}p{1.2cm}p{1.2cm}}
100 GHz&143 GHz&217 GHz&353 GHz&545 GHz\\
\multicolumn{5}{c}{\includegraphics[width=8cm]{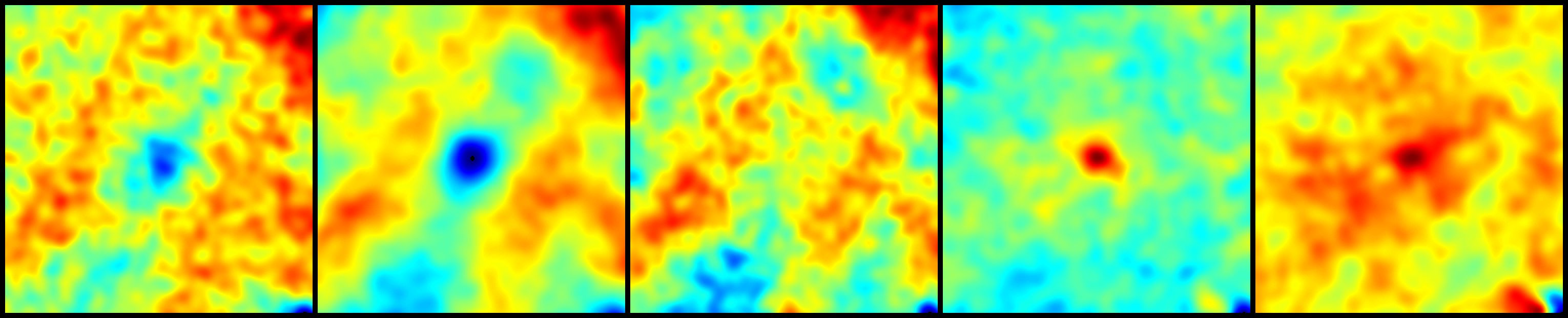}} \\[-0.14cm]
\multicolumn{5}{c}{\includegraphics[width=8cm]{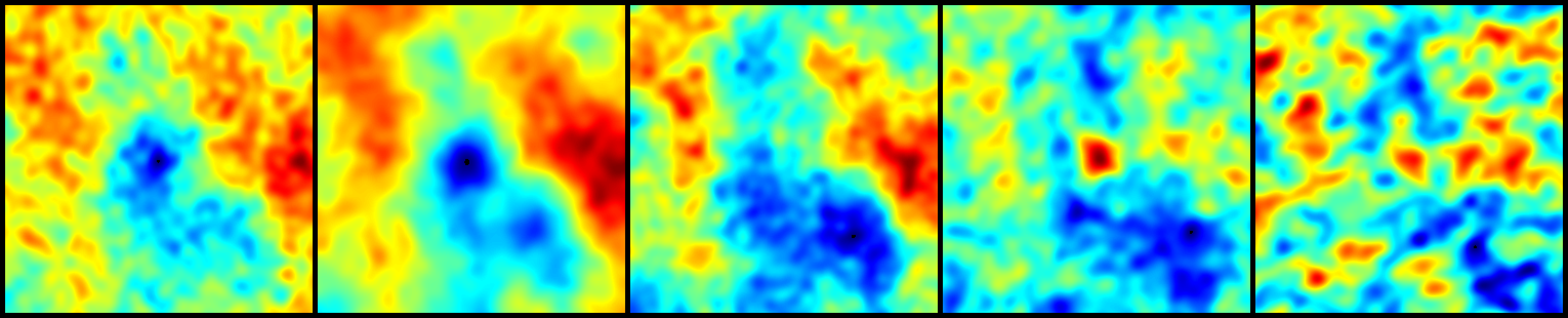}} \\[-0.14cm]
\multicolumn{5}{c}{\includegraphics[width=8cm]{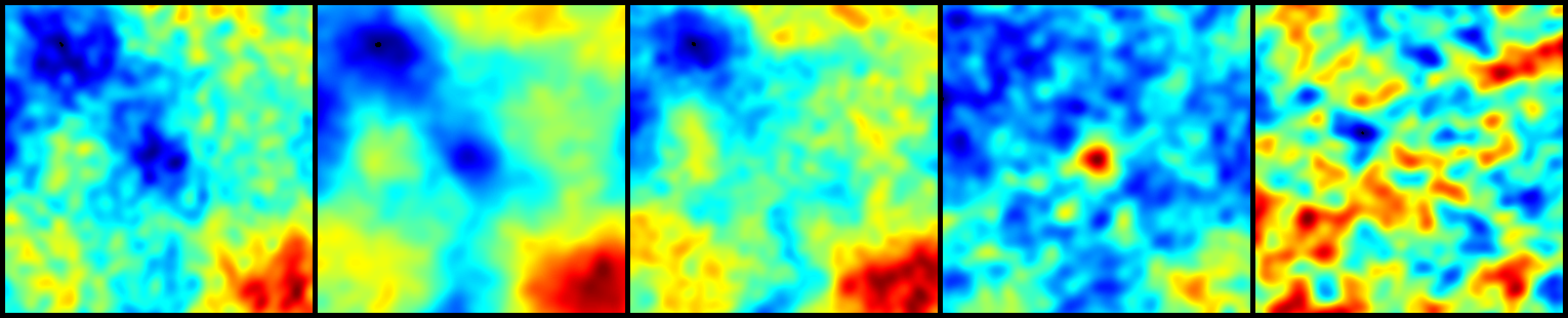}} \\[-0.14cm]
\multicolumn{5}{c}{\includegraphics[width=8cm]{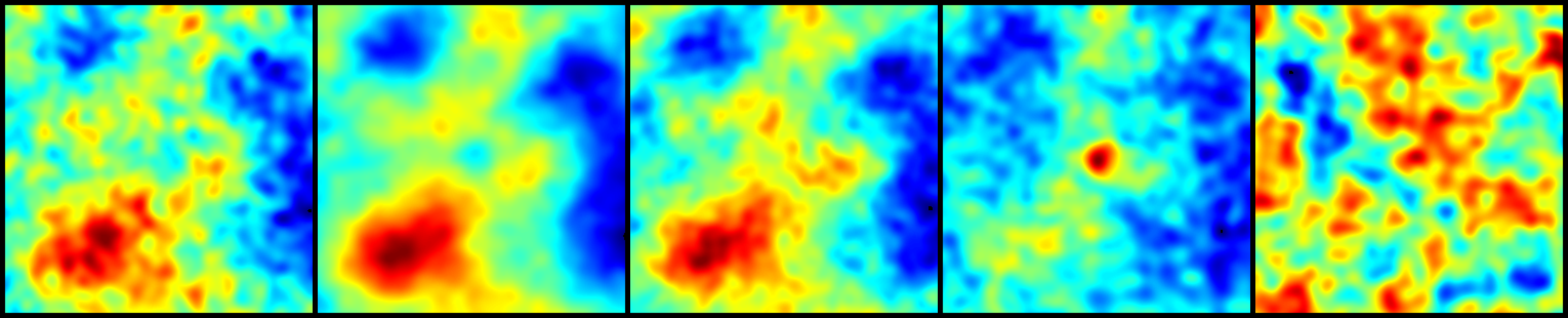}} \\[-0.14cm]
\multicolumn{5}{c}{\includegraphics[width=8cm]{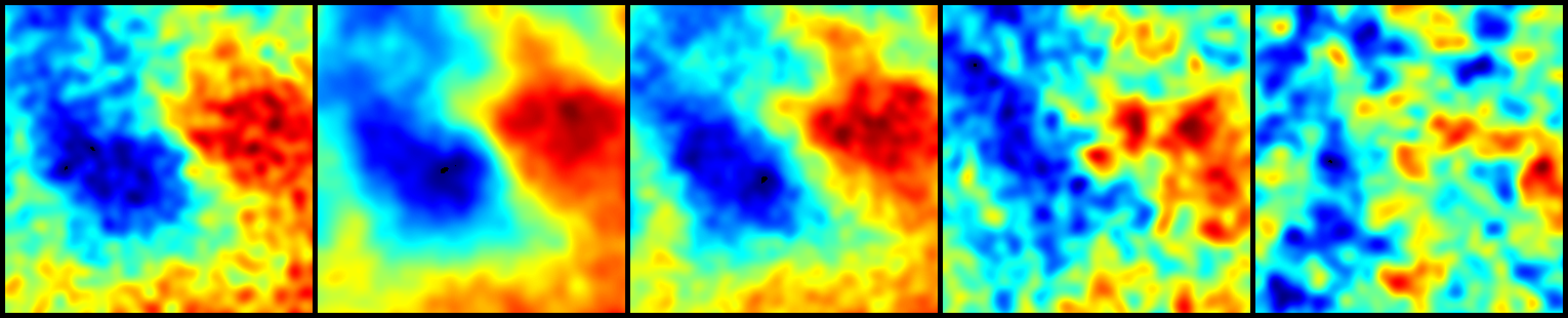}} \\[-0.14cm]
\multicolumn{5}{c}{\includegraphics[width=8cm]{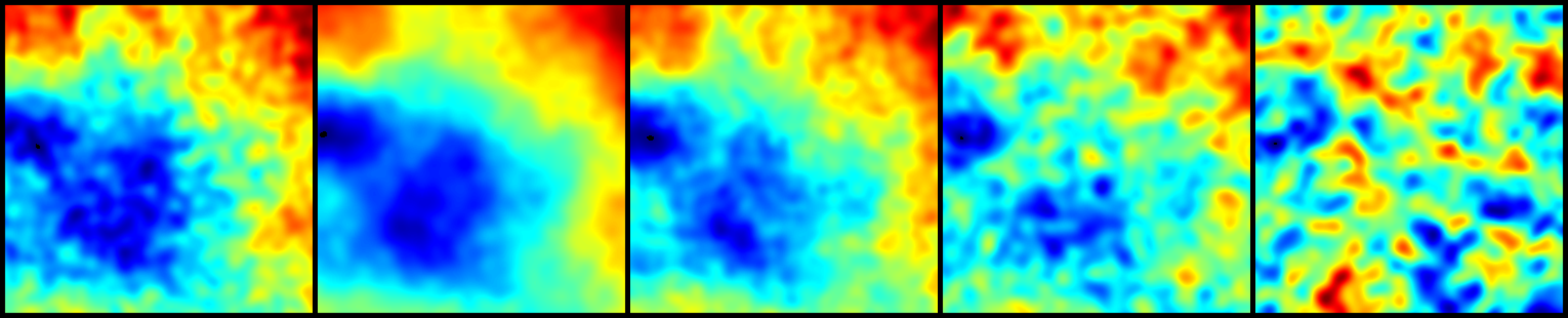}} \\[-0.14cm]
\multicolumn{5}{c}{}\\[-0.14cm]
\multicolumn{5}{c}{\includegraphics[width=8cm]{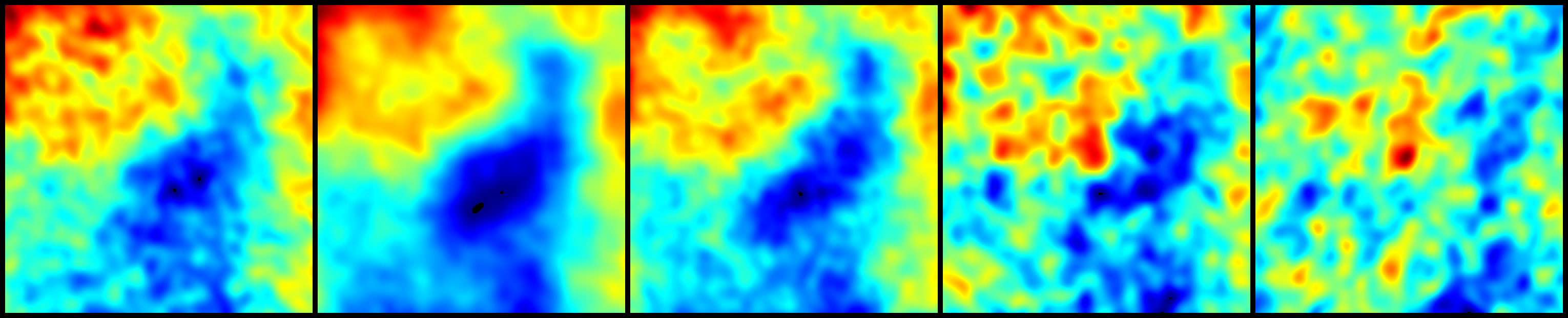}} \\[-0.14cm]
\multicolumn{5}{c}{}\\[-0.14cm]
\multicolumn{5}{c}{\includegraphics[width=8cm]{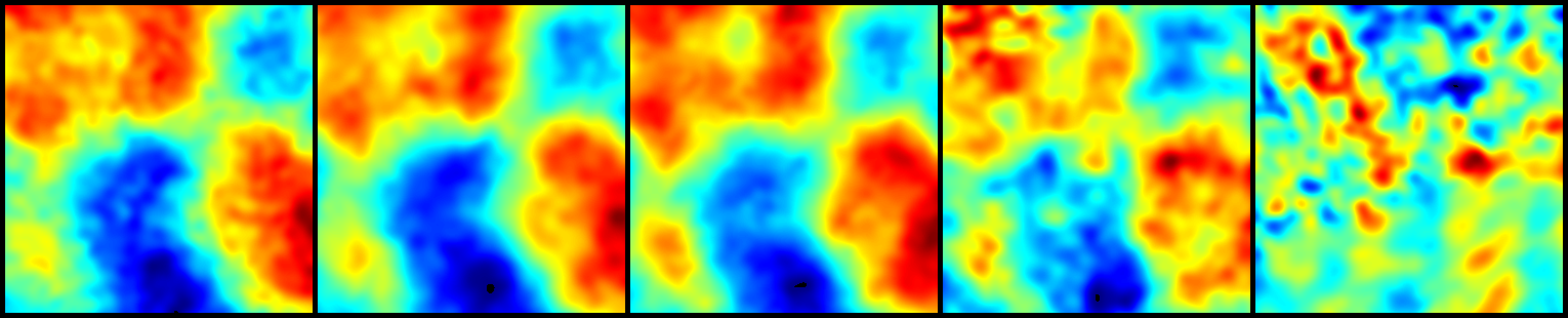}} \\[-0.14cm]
\end{tabular}
\caption{Same as Fig.~\ref{stack1}, but for redshift bins from z=0.5 to z=1.0.}
\label{stack2}
\end{figure}

\section{Methodology}
\label{sec_meth}
In this section, we describe the procedure that was used to extract
the tSZ signal from the full-sky {\it Planck} intensity maps.  We
start by considering clusters from the {\it Planck} catalog, because they have a
high signal-to-noise ratio and a low level of contamination. Then, we
discuss the case of additional clusters from the samples listed in
Sect.~\ref{sec_cat}.\\
In a first step, we set all {\it Planck} frequency channels to a
common angular resolution of 10', corresponding roughly to the angular
resolution the lowest resolution channel at 100~GHz.  Then, we
extracted individual patches centered on the location of {\it Planck}
galaxy clusters and stacked them.  Next, we cleaned the obtained
stacked patches from IR emission from galactic dust and extragalactic
IR galaxies using the 857~GHz channel. Finally, we estimated the tSZ
flux in each cleaned stacked patch.

\subsection{Stacking}
\label{secstack}
To increase the significance of the tSZ signal per frequency,
we performed a stacking analysis in multiple redshift bins.  Our
sample consists of 813 clusters with known redshifts, distributed in
20 redshift bins from $z=0$ to $z=1$, with a binsize $\Delta z = 0.05$.
The redshift bins and number of clusters per bin are listed in
Table.~\ref{tabbinrad}. Note that the mean redshift $\bar{z}$ and the
weighted mean over $Y_{500}$, $z_{\rm eff}$ only differs at the third
decimal. This shows that the SZ flux distribution over the bin is
homogeneous and that the choice of bins is a fair sampling of the
redshift distribution for the considered cluster sample. In the
following, we use $z_{\rm eff}$ as the mean value for our redshift
bin.

We constructed a stacked patch per redshift bin. We thus extracted
individual patches of 2$^\circ \times$2$^\circ$\footnote{We chose a
  constant patch size because the majority of clusters are point-like with
  respect to the adopted resolution of $10'$.  Differences in the
  physical extension of clusters have thus no significant impact on
  our results.} from the full-sky intensity maps with pixels of
1.7'. The individual patches were centered on the positions of the
clusters in the considered redshift bin. The individual patches were
re-projected using a nearest neighbor interpolation on a grid of 0.2'
pixels in gnomonic projection. This conserves the cluster flux. Note
that re-projection effects are the same for all frequencies and
consequently do not bias the estimation of $T_{\rm CMB}$ since the
latter only depends on the shape of the tSZ spectral law.\\
For each redshift bin individual patches are cleaned from point-source
contamination. To do this, we masked all sources from the {\it Planck} Catalog of Compact
Sources \citep{PlanckCCS} detected above 10 $\sigma$ in one of the six
highest frequency channels of {\it Planck} within an area of 30' from the
cluster center. Sources within a radius of
20' from the considered cluster were not masked. This process avoids
biases in the dust-cleaning process (see Sect.~\ref{dclean}).\\
Finally in each redshift bin, the individual patches per frequency were
co-added with a constant weight. This choice is a two-fold one. It
accounts for the fact that the main contribution to the noise
due to CMB is similar from one patch to the other. Furthermore, it avoids
cases where a particular cluster will dominate the stacked signal. It
is worth noting that this a choice is not optimized for the
signal-to-noise ratio of the flux.

We verified that we derive similar results when applying random rotations
to the extracted patches before stacking. This test demonstrated that
our stacking is not sensitive to specific orientations, produced, for
example, by the thermal dust emission from the Galactic plane.

\subsection{Dust cleaning}
\label{dclean}
To remove thermal dust contamination in stacked patches per frequency,
we used the 857~GHz channel. The thermal dust emission is known to have a varying spectral energy distribution (SED) across the sky \citep{PlanckDUST2}.
To compute the effective thermal dust SED for each stacked patch per frequency, we assumed that in a field of 2$^\circ \times$2$^\circ$ the thermal dust spectral properties can be
well described by a single SED.\\

This cleaning was performed for each redshift bin $\delta$ and for
each frequency $i$ by computing the scale factor, $\rho^\delta_i$,
between the stacked patch at 857~GHz, $M^\delta_{857}$, and other
stacked patches from 100 to 545~GHz, $M^\delta_{i}$. To avoid
biases produced by correlations between IR emission and the tSZ
effect, we computed $\rho^\delta_i$ by removing the central region of
$r<30'$, with $r$ the angular distance to the clusters position.
We checked that performing the cleaning per cluster instead of per redshift bin provides comparable results.
\begin{equation}
\rho^\delta_i = \frac{\left(\sum_{r>30'} M^\delta_{857}M_{i}\right)/n
  - \left(\sum_{r>30'}
  M^\delta_{857}\right)\left(\sum_{r>30'}M^\delta_{i}\right)/n^2
}{\left(\sum_{r>30'} (M^\delta_{857})^2\right)/n - \left(\sum_{r>30'}
  M^\delta_{857}\right)^2/n^2},
\end{equation}
where $n$ is the number of pixels satisfying the condition
$r>30'$. 
Then, the stacked patches cleaned from thermal dust were
obtained by subtracting the 857~GHz from the other channels:
\begin{equation}
M_i^{\delta,{\rm cleaned}} = M^\delta_{i} - \rho^\delta_i M^\delta_{857}.
\end{equation}
This cleaning process has an impact solely on 353 and 545 GHz channels with $\rho^\delta_i$ values varying only by 10\% and 5\% at 353 and 545 GHz, from bin to bin.
Dust-cleaned stacked patches for each redshift bin are presented in
Figs.~\ref{stack1}~and~\ref{stack2}. We clearly observe \citep[as
  reported in][]{PlanckSZC} the intensity decrement at low frequency
($< 217$~GHz), the null effect at 217~GHz and the positive emission at
high frequency ($> 217$~GHz). Note that we report a clear detection of
the tSZ effect even in the 545~GHz {\it Planck} channel for redshifts
up to 0.6.

\subsection{Flux estimate}
\label{fluxest}
To measure the flux from the cleaned stacked patches at 100, 143, 217,
353, and 545~GHz, we first derived the shape of the tSZ signal in the
stacked patches.  To do this, we computed a Compton-$y$ parameter map for each cluster
individually using the MILCA method
\citep{hur13}. Computing the $y$-map on a stacked patch would lead to a
poorer tSZ reconstruction. We then stacked the reconstructed $y$-maps
for the clusters within a given redshift bin. We refer to them as
tSZ filter or template.\\
The tSZ signal in the stacked patches presents a circular symmetry. We
therefore computed the radial profile of the tSZ filter and re-projected it
to construct a denoised tSZ filter, $M^\delta_{\rm tSZ}$.  We
used $M^\delta_{\rm tSZ}$ as a shape filter to measure the tSZ flux,
$F^\delta_i$, in each cleaned stacked patch $M^{\delta,{\rm
    cleaned}}_{i}$. \\ 
The tSZ flux was obtained by computing a linear fit of the denoised tSZ
filter on the cleaned stacked patches per frequency in a radius of 20'
around the cluster center, assuming homogeneous noise and the
following modeling
\begin{equation}
M^{\delta,{\rm cleaned}}_{i} = {F}^\delta_{i} M^\delta_{\rm tSZ} + b^\delta_i + N^\delta_i,
\label{specest}
\end{equation} 
with $b^\delta_i$ a constant baseline accounting for large-scale ($>20'$) residual contamination and $N^\delta_i$ the noise
component including astrophysical emissions and instrumental
noise.  

\begin{figure*}[!th]
\begin{center}
\includegraphics[width=17.5cm,trim = 7cm 7.5cm 8cm 8cm, clip]{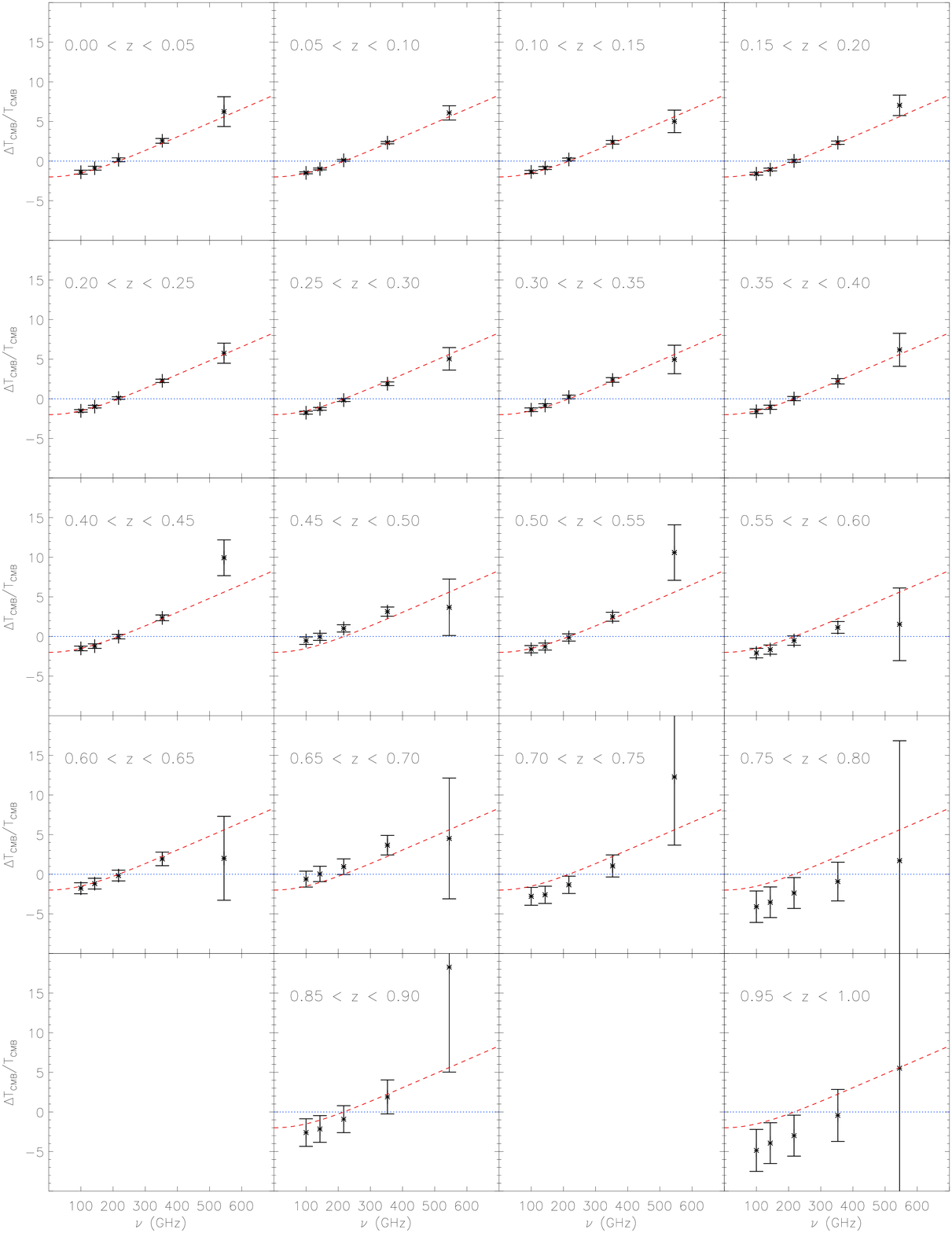}
\caption{From left to right and top to bottom: Measured tSZ emission
  law in units of $K_{\rm CMB}$ for each redshift bin form 0 to 1,
  with $\Delta z=0.05$. Note that bins with $0.80 \leq z<0.85$ and
  $0.90 \leq z<0.95$ do not contain any cluster. Black stars denote
  the data and dashed red lines the theoretical tSZ emission
  law. Error bars are strongly correlated between the
  different frequency channels, see Tab.~\ref{covmat}.}
\label{spec}
\end{center}
\end{figure*}

From Eq.~\ref{specest}, we derived a tSZ emission law,
$\widehat{F}^\delta_{i}$, for each redshift bin $\delta$. The
estimator $\widehat{F}^\delta_{i}$ of ${F}^\delta_{i}$ was obtained
by adjusting both ${F}^\delta_{i}$ and $b^\delta_i$.
Figure~\ref{spec} presents the derived emission law for the tSZ effect
for each redshift bin. The standard model is displayed as a dashed red
line. Measured fluxes, $\widehat{F}^\delta_{i}$, were used to
derive the value of $T^\delta_{\rm CMB}$ for each redshift bin in
Sect.~\ref{sec_ana}.

Note that using of a tSZ template induces no prior on the tSZ
spectral law (considering that the spatial distribution of the tSZ
signal is the same in all channels) and thus on
$\widehat{F}^\delta_{i}$.  The small difference between tSZ filters
from bin to bin is due to the difference in cluster extensions since
the profiles used here were not rescaled with respect to their angular sizes contrary to \citet{arn10} and the \citet{PlanckPPP}.

\section{Statistical and systematic uncertainties}
\label{sec_error}
We now discuss the sources of uncertainties and systematic errors on
the measured fluxes $\widehat{F}^\delta_{i}$. We first focus on the
uncertainties produced by background and foreground signals, including
instrumental noise. Then we address the astrophysical components
correlated to the tSZ emission, and finally we discuss the systematic
errors caused by the instrument spectral responses.

\subsection{Background and foreground contamination}
\label{errstat}
To estimate the uncertainty on $\widehat{F}^\delta_{i}$ caused by
contamination by other sources signal, we extracted the flux
at 1000 random positions across the sky for each denoised tSZ
filter. We avoided the galactic plane area, which is not represented in
our cluster distribution.

In the stacking process each cluster was considered to be uncorrelated with
others. Consequently, we derived the full covariance matrix for the
flux estimate in frequency channels from 100 to 545~GHz. An example of correlation matrix is presented in Table.~\ref{covmat}. This
correlation matrix only accounts for uncertainties produced by
uncorrelated components (with respect to the tSZ effect) in the flux
estimation. This correlation matrix has a determinant of $10^{-4}$,
which quantifies the volume occupied by the swarm of data samples in
our five-dimension subspace (cleaned frequencies from 100 to
545~GHz).\\
At 100 to 217~GHz, the CMB anisotropies are the main source of
uncertainties. This explains the high level of correlation in the
estimated fluxes. At higher frequencies, instrumental noise and
dust residuals become an important contribution to the total
uncertainties, which explains the lower level of correlation.  

\begin{table}
\caption{Correlation matrix of statistical uncertainties for the
  measurement of tSZ flux $\widehat{F}^\delta_{i}$, estimated from
  1000 random positions across the sky.}
\label{covmat}
\center
\begin{tabular}{|c|c|c|c|c|c|}
\hline
Frequency (GHz) & 100 & 143 & 217 & 353 & 545 \\ \hline
100 &1  &   0.97    &  0.96   &   0.79    & 0.23 \\ \hline
143 &0.97    &   1    &  0.99   &   0.83     & 0.26 \\ \hline
217 & 0.96     & 0.99   &    1   & 0.86     & 0.32 \\ \hline
353 &  0.79      &0.83  &   0.86   &   1   &   0.61 \\ \hline
545 &    0.23     &0.26 &     0.32   &  0.61  &     1 \\ \hline
\end{tabular}
\end{table}

\subsection{Correlated foreground contamination}

Uncertainties produced by uncorrelated foregrounds, for instance, CMB, can be
fairly estimated from measurements at random positions over the
sky. However, these measurements do not account for noise or
bias produced by correlated emissions, such as, radio sources at
low frequency, cosmic infrared background (CIB) at high frequency,
and the kinetic SZ effect.

\subsubsection{Cosmic infrared background}
\label{seccib}
The CMB temperature measure is directly related to the frequency at
which tSZ effect is null, making it a key frequency in our
analysis. At 217~GHz, a contribution from the CIB (the integrated IR
emission from dust in distant galaxies \citep[see e.g.,][for
  reviews]{hau01,kas05,lag05}) is significant and is correlated to the
tSZ signal \citep[e.g., ][]{add12}. In the following, we assumed a
conservative situation of full correlation between tSZ and CIB,
$\rho^\delta_{\rm cor} = 1$. Note that the cluster sample used here
does not contain all clusters at $z<1$. Consequently, the actual
correlation factor $\rho^\delta_{\rm cor}$ is lower than one and
depends on the completeness with respect to the redshift
\citep[see][for more details]{PlanckSZC}.

Considering the CIB intensity at 217 GHz, $\ell C^{\rm CIB}_\ell =
0.25$~$\mu$K$^2$.sr, given by the \citet{PlanckCIB}, the CMB power
spectrum at $\ell=1000$, $\ell C^{\rm CMB}_{\ell} \simeq 1
$~$\mu$K$^2$.sr, given in \citet{PlanckPS} and considering the
contribution to tSZ power spectrum from our sample $\ell C^{\rm
  tSZ}_\ell \simeq 0.2~10^{-15}$~$y^2$.sr \citep{PlanckSZS}, we can
compute the bias due to the CIB contribution. We furthermore assumed that 90\%
of the CIB is cleaned by the dust-cleaning process discussed in
Sect.~\ref{dclean} and we defined $f_{clean} = 0.9$.

We used the following equation for the bias induced by tSZ$\times$CIB
correlation for a given redshift bin $\delta$:
\begin{equation}
B^\delta_{\rm CIB} = \rho^\delta_{\rm cor} \frac{f_{\rm CIB}}{N_{\rm
    bin}} \sqrt{\frac{C^{\rm CIB}_\ell}{g(\nu_i)^2C^{\rm tSZ}_\ell}}
\frac{F^{\delta}_i}{\left( \Delta F^{\delta} \right)_{\rm CMB}}
\left(1-f_{\rm clean} \right) ,
\end{equation}
$f_{\rm CIB} \sim 0.05$ is the fraction of CIB emission produced
by objects at redshift $<1$, $N_{\rm bin}$ is the number of redshift
bins, and $\left( \Delta F^{\delta} \right)_{\rm CMB}$ is the
contribution of the CMB to the uncertainties over $F_i^{\delta}$. 

We derived a ratio between CMB and CIB fluctuations of about 0.2\%
on average per cluster.  At most, we use about one hundred clusters in
a single redshift bin. This reduced the CMB fluctuations
by a factor 10. Consequently, in a conservative case, the CIB at 217~GHz
contributes at the level of 2\% of the intensity of CMB and can therefore
be neglected.

\subsubsection{Radio point sources}
\label{radconta}
Radio sources within galaxy clusters can produce an overestimate of
the flux at low frequencies, which in turn leads to an underestimate
of $T_{\rm CMB}(z)$.

Using the NVSS and the SUMSS catalogs of radio point sources, we
estimated the radio point-source contamination on the measured tSZ
flux. To do this, we projected the NVSS and the SUMSS catalogs on a
full-sky map (considering only SUMSS sources at Dec$~<-40^\circ$, and
extrapolating their fluxes from 853~MHz to 1.4 GHz with a spectral
index of -1). Then, we smoothed the obtained map
at 10'.  We thus obtained a full-sky map of the combined radio
sources at 1.4~GHz on which we estimate the radio flux $F^\delta_{\rm rad}$ for each stacked patch and
redshift bin $\delta$ using the
approach described in Sect.~\ref{fluxest}. Finally, we extrapolated the
radio emission from 1.4~GHz to {\it Planck} frequencies, assuming a
spectral index of -1. We estimated the spectral index by
computing the cross power spectrum between the NVSS catalog projected
map and the {\it Planck} 100~GHz channel. We derived an averaged
spectral index of $-0.995 \pm 0.010$.

The radio fluxes within the tSZ filter, $F^\delta_{\rm rad}$,
expressed in terms of percentage of the measured tSZ flux at 100~GHz
are summarized in Table~\ref{tabbinrad}. Under the simple assumption of a
single spectral index -1, we show that the contamination by
radio source emission on the measured tSZ flux at 100 GHz is lower than
15\%. In the analysis described in Sect.~\ref{fitconta}, we show
how this contribution was modeled, fitted, and accounted for in the
estimate of the uncertainties.

\begin{table}
\center
\caption{Reshift binning used for our analysis, $N_{cl}$ is the number of clusters per bin, $\bar{z}$ is the mean redshift, $z_{\rm eff}$ is the mean redshift weighted by the flux of each clusters $Y_{500}$ \citep{PlanckSZC} and $F^\delta_{\rm sync}$is the radio contamination in tSZ measured flux at 100 GHz estimated from NVSS and SUMSS data. It is expressed in terms of percentage of the tSZ flux at 100~GHz.}
\label{tabbinrad}
\begin{tabular}{|c|c|c|c|c|}
\hline
Redshift bin $\delta$ & $N_{cl}$ & $\bar{z}$ & $z_{\rm eff}$ & $F^\delta_{\rm sync}$ (\%)\\ \hline
0.00-0.05 & \phantom{0}43 & 0.036 & 0.037&\phantom{-}14.0\\
0.05-0.10 & 125 & 0.076 & 0.072&\phantom{-}10.3\\
0.10-0.15 & \phantom{0}92 & 0.126 & 0.125&\phantom{-0}6.2\\
0.15-0.20 & 104 & 0.172 & 0.171&\phantom{-0}6.0\\
0.20-0.25 & \phantom{0}95 & 0.221 & 0.220&\phantom{-0}4.5\\
0.25-0.30 & \phantom{0}87 & 0.273 & 0.273&\phantom{-0}9.0\\
0.30-0.35 & \phantom{0}81 & 0.323 & 0.322&\phantom{-0}1.0\\
0.35-0.40 & \phantom{0}50 & 0.375 & 0.377&\phantom{-0}4.9\\
0.40-0.45 & \phantom{0}45 & 0.424 & 0.428&\phantom{-0}0.4\\
0.45-0.50 & \phantom{0}26 & 0.471 & 0.471&\phantom{-0}4.8\\
0.50-0.55 & \phantom{0}20 & 0.525 & 0.525&\phantom{-0}2.5\\
0.55-0.60 & \phantom{0}18 & 0.565 & 0.565&\phantom{0}-0.5\\
0.60-0.65 & \phantom{0}12 & 0.619 & 0.618&\phantom{-0}2.5\\
0.65-0.70 & \phantom{00}6 & 0.676 & 0.676&\phantom{0}-3.2\\
0.70-0.75 & \phantom{00}5 & 0.718 & 0.718&\phantom{-0}3.7\\
0.75-0.80 & \phantom{00}2 & 0.783 & 0.777&\phantom{0}-2.8\\
0.80-0.85 & \phantom{00}0 & --- & ---&---\\
0.85-0.90 & \phantom{00}1 & 0.870 & 0.870&\phantom{0}-0.3\\
0.90-0.95 & \phantom{00}0 & --- & ---&---\\
0.95-1.00 & \phantom{00}1 & 0.972 & 0.972&\phantom{0}-0.3\\
\hline
\end{tabular}
\end{table}

\subsubsection{Kinetic Sunyaev-Zel'dovich effect}
The kinetic SZ effect, kSZ, is the Doppler shift of CMB photons that
scatter the intracluster electrons. This effect is faint, one order
of magnitude lower than the tSZ. It has the same spectral dependance as
the CMB and is spatially correlated to the tSZ signal. The kSZ
effect can produce positive or negative CMB temperature anisotropies.
Consequently, this effect will not bias the tSZ measurement,
but will enlarge the CMB dispersion at the clusters position and
therefore the error-bars. At {\it Planck} resolution, the increase in
CMB variance due to the kSZ is small \citep{PlanckkSZ} and can be
neglected in our analysis.\\
In some nonstandard inhomogenous cosmological models \citep[see
  e.g.][]{goo95,cla12}, the kSZ monopole is different from zero. This
could induce a bias in the tSZ measurement. However, these models are
now strongly constrained by {\it Planck} data \citep{PlanckkSZ}.

\subsection{Effects of bandpass and calibration uncertainties}

To measure the tSZ emission law in the {\it Planck} channels we
integrated the tSZ emission law, $g(T^\delta_{\rm CMB},T^\delta_{\rm
  e},\nu)$ (see Eq.~\ref{szspec}), over the {\it Planck} spectral
responses (i.e., bandpasses), see \citet{PlanckBP}, $H_i$, in the
following manner:
\begin{equation}
\centering A_i(T^\delta_{\rm CMB}, T^\delta_{\rm e})= \frac{\int
  H_i(\nu)g(T^\delta_{\rm CMB},T^\delta_{\rm e},\nu) {\rm d}\nu}{\int
  H_i(\nu)C(\nu){\rm d}\nu},
\label{sztrans}
\end{equation}
with, $A_i$ the tSZ transmission in the $i$-th {\it Planck} channel,
$C(\nu)$ the emission law of the calibrators, CMB, and planets
\citep{PlanckCAL}, and $T^\delta_{\rm e}$ the effective cluster
temperature per redshift bin. 

The bandpasses present uncertainties that depend on the tSZ spectral
law. Given that $T^\delta_{\rm CMB}$ and $T^\delta_{\rm e}$ produce
small variations of the tSZ, we can assume that the bandpass uncertainties
are constant.  We propagated the uncertainties on the bandpasses $H_i$
in our analysis by computing the uncertainties on the tSZ emission law
$A_i(T_{\rm CMB}, T^\delta_{\rm e})$ after integrating over the
bandpass. These turn into uncertainties on the recovered value of
$T^\delta_{\rm CMB}$. We derived a systematic uncertainty of $0.010$~K
at $z=0$ for $T^\delta_{\rm CMB}$; it becomes $0.010~\times~(1+z_{\rm
  eff})$~K for each redshift bin.

At low frequencies, the {\it Planck} data were calibrated with respect
to CMB dipole ($T_{\rm CMB}(z=0)$). Uncertainties on the calibration
can also lead to bias in $T^\delta_{\rm CMB}(z)$. However, this source
of uncertainties is small, about 0.2\% at 100 to 217~GHz
channels \citep{PlanckCAL}. It was neglected in the following.

\section{Analysis}
\label{sec_ana}

\subsection{Modeling the signal}

Our measurement at the i-th {\it Planck} frequency can be modeled as
\begin{equation}
F^\delta_i = Y^\delta A_i(T^\delta_{\rm CMB},T^\delta_{\rm e}) +
F^\delta_{\rm rad}A_i^{\rm rad} + F^\delta_{\rm ir}A_i^{\rm ir},
\label{specmod}
\end{equation}
with $Y^\delta~=~\int_\delta\int y {\rm d}\Omega{\rm d}z$ the
integrated Compton parameter for a redshift bin $\delta$, $A_i^{\rm
  rad}$ the radio source spectrum with spectral index -1 normalized
to 1 at 100~GHz, and $A_i^{\rm ir}$ is the IR spectrum with dust
temperature $T_{\rm d} = 17$K and spectral index $\beta_{\rm d} = 1.8$ \citep[][]{PlanckDUST}
normalized to 1 at 353~GHz. We verified that this $A_i^{\rm ir}$ SED theoretical model is consistent with the spectrum derived from our cleaning procedure.

The adjustable parameters are ${Y^\delta}$, $T^\delta_{\rm CMB}$,
$T^\delta_{\rm e}$, $F^\delta_{\rm rad}$ the radio source flux and
$F^\delta_{\rm ir}$ the IR contamination level including CIB emission.

\begin{figure}[h!]
\center
\includegraphics[width=8cm,trim=4cm 0cm 0cm 0cm]{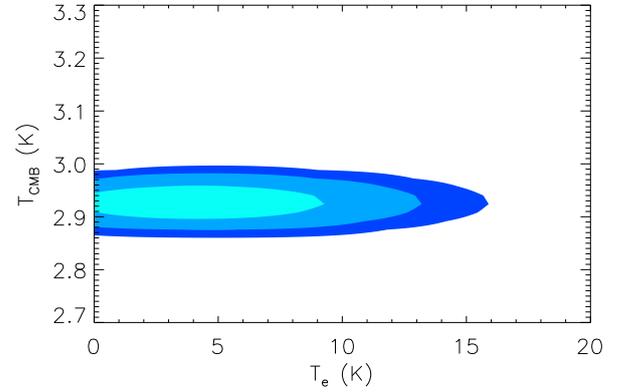}
\caption{Likelihood function of the measured tSZ emission law as a
  function of $T^\delta_{\rm CMB}$ and $T^\delta_{\rm e}$ for redshift
  bins between $z=0.05$ and $z=0.10$. Confidence levels at 1, 2, and
  $3\sigma$ are presented in light-blue, blue, and dark-blue areas.}
\label{tecor}
\end{figure}
\subsection{Sensitivity analysis}
In the following, we perform a sensitivity analysis between
$T^\delta_{\rm CMB}$ and the other parameters of the model.

\subsubsection{Impact of relativistic corrections}
The relativistic corrections to the tSZ emission law has been computed
numerically \citep[see e.g.][]{rep95,poi98,ito98}.  If we assume that
the relativistic corrections can be described as a first-order
approximation,
\begin{equation}
\Delta T^{\rm relat}_{\rm CMB}(T_{\rm e}) = \Delta T^{\rm
  unrelat}_{\rm CMB} + T^\delta_{\rm e} \Delta T^{\rm cor}_{\rm CMB},
\end{equation}
the averaged tSZ emission law can be described with an averaged
temperature fitted as an effective temperature, $T^\delta_{\rm e}$,
for the stacked tSZ signal \citep[see][for a more detailed fitting
  formula]{ito00}.

Figure~\ref{tecor} presents the likelihood function in the plane
($T^\delta_{\rm CMB}$, $T^\delta_{\rm e}$) for clusters in the
redshift bin 0.05 to 0.10. The other redshift bins present similar
behaviors. The value we derived for $T^\delta_{\rm e}$ is below 10 keV
at $1\sigma$ level and below 15 keV at $3\sigma$ level, with a
best-fitting value around 5 keV. This is consistent with the X-ray
derived temperatures for typical clusters in the {\it Planck} sample.
Figure~\ref{tecor} shows no correlation between the recovered value of
$T^\delta_{\rm CMB}$ and $T^\delta_{\rm e}$. This Indicates that we can
safely neglect relativistic corrections in our analysis.

\subsubsection{Radio and IR contaminations}
\label{fitconta}

\begin{figure}[h!]
\begin{center}
\includegraphics[width=8cm,trim=4cm 0cm 0cm 0cm]{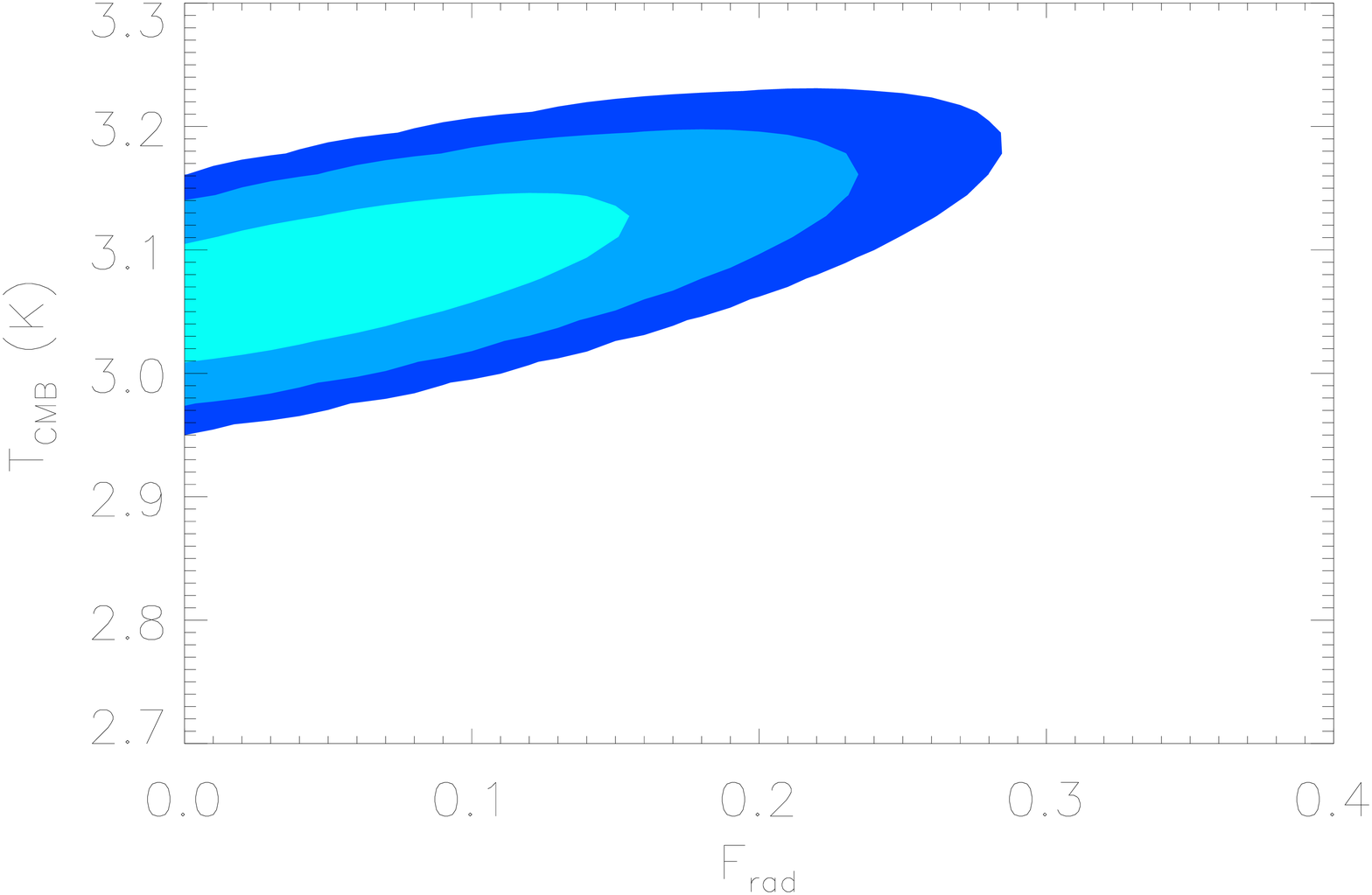}\\
\includegraphics[width=8cm,trim=4cm 0cm 0cm 0cm]{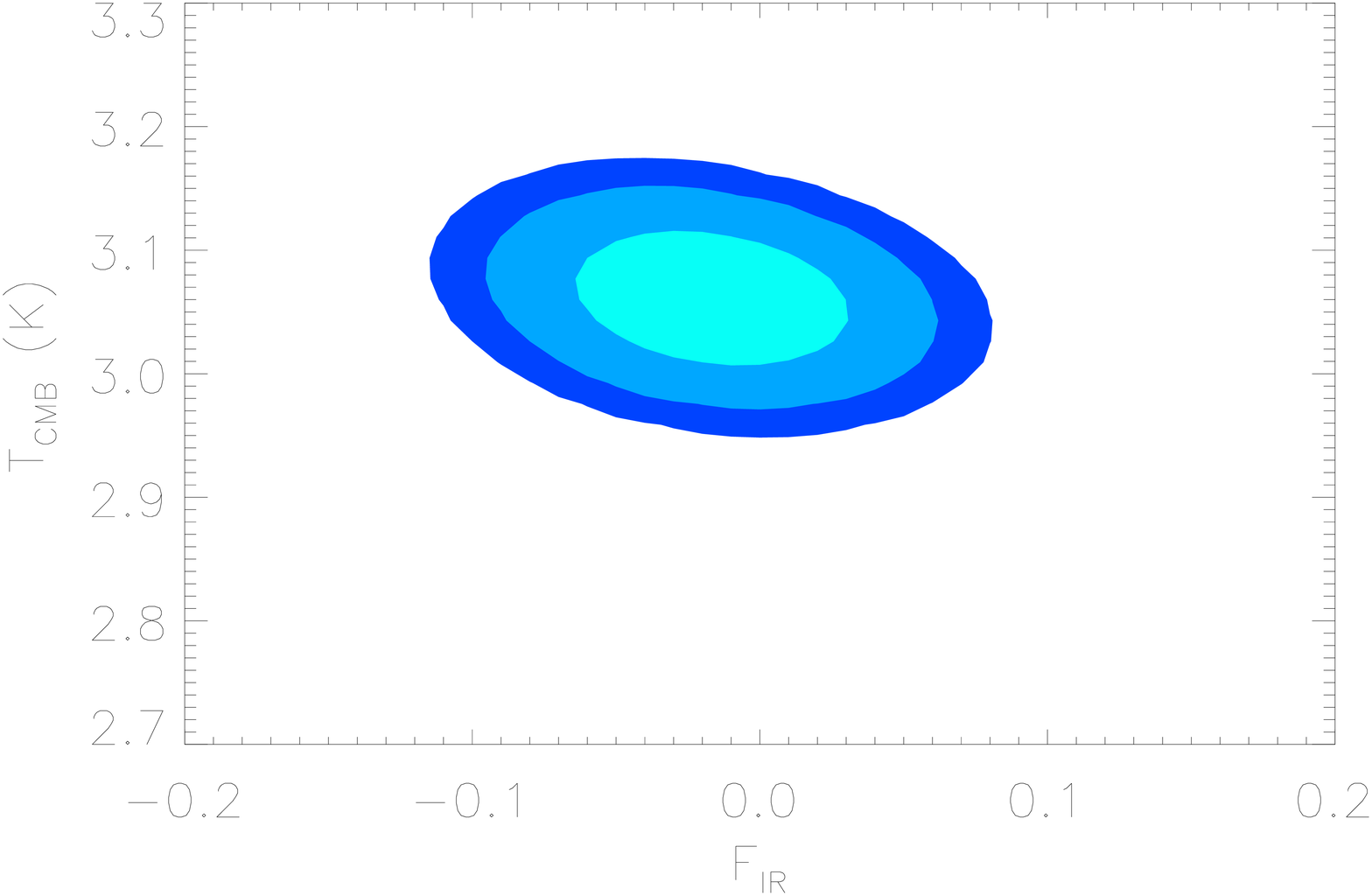}
\caption{Top panel: Likelihood function of the measured tSZ emission
  law as a function of $T^\delta_{\rm CMB}$ and the radio source flux,
  expressed in units of tSZ flux at 100~GHz, for redshifts between
  $z=0.10$ and $z=0.15$. Bottom panel: Likelihood function of the
  measured tSZ emission law as a function of $T^\delta_{\rm CMB}$ and
  the infrared contamination level, expressed in units of tSZ flux at
  353~GHz, for redshift between $z=0.10$ and $z=0.15$. Confidence
  levels at 1, 2, and $3\sigma$ are represented in light-blue, blue, and
  dark-blue.}
\end{center}
\label{ircor}
\end{figure}

Figure~\ref{ircor} presents the likelihood function in the plane of
the radio contamination, $F^\delta_{\rm srad}$, and $T^\delta_{\rm
  CMB}$ (top panel), and in the plane of the IR contamination,
$F^\delta_{\rm IR}$, and $T^\delta_{\rm CMB}$ (bottom panel) for
clusters within a redshift bin 0.10 to 0.15. The other redshift bins
present similar behaviors.  We found that the IR contamination,
including CIB residuals, is compatible with 0 and does not bias our
estimate of the tSZ flux. We also note that the derived value of $F^\delta_{\rm IR}$ is consistent with all $f_{\rm clean} > 0.8$ at 6 $\sigma$ level, consistent with the conservative case discussed in Sect.~\ref{seccib}.\\
Furthermore, we observed that the radio source contamination produces 
some bias on the $T^\delta_{\rm CMB}$ measurement. This 
bias translates into $\widehat{T^\delta}_{\rm CMB} = T^\delta_{\rm CMB} - 0.005
\times (1+z_{\rm eff}) \times F^\delta_{\rm rad}$.

\subsection{Profile likelihood analysis}

Given the results of the sensitivity analysis discussed above, we can
simplify the model in Eq.~\ref{specmod} to
\begin{equation}
F^\delta_i = {Y^\delta}A_i(T^\delta_{\rm CMB}) + F^\delta_{\rm rad}A_i^{\rm rad},
\label{specsim}
\end{equation}
considering only the relevant parameters ${Y^\delta}$,
$T^\delta_{\rm CMB}$ and $F^\delta_{\rm rad}$.

To fit $T^\delta_{\rm CMB}$ in each redshift bin, we used a
profile likelihood approach. First, we computed, through an unbiased
linear fit, the tSZ flux, $\widehat{Y}^\delta$, of our measured
spectral law for each value of $T^\delta_{\rm CMB}$ and $F^\delta_{\rm
  rad}$. Given the similar amplitudes of the uncertainties on the
measurement (mainly CMB contamination) and the model (bandpasses), we
used the following estimator for ${Y^\delta}$:
\begin{equation}
\widehat{Y}^\delta = \left[ {\bold A}^{T}{\cal W}{\bold A} -
  {\mathrm{Tr}({\cal C}_{A}^{T}{\cal W}}) \right]^{-1}\left[ {\bold
    A}^{T}{\cal W}\widehat{\bold F}^\delta \right],
\end{equation}
with ${\bold A}$ the tSZ transmission vector defined in
Eq.~\ref{sztrans}, ${\cal C}_{A}$ the ${\bold A}$ covariance
matrix\footnote{The uncertainties on the response $A_i^{\rm rad}$ are
  lower than 1\% and neglected. In contrast, the uncertainties on
  the response for the tSZ in the 217~GHz channel is about 25\%
  \citep{PlanckBP}.}, $\widehat{\bold F}^\delta$ the measured tSZ
emission law (see Eq.~\ref{specest}), and ${\cal W} = {\cal
  C}^{-1}_{F^\delta}$ the inverse of the noise covariance matrix on
$\widehat{\bold F}^\delta$.  

Then, we computed the $\chi^2$, for each paire of parameters
($T^\delta_{\rm CMB}$, $F^\delta_{\rm sync}$) as
\begin{equation}
\chi^2 = \left(\widehat{\bold F}^\delta - {\bold F}^\delta \right)^T
\left[{\cal C}_{F^\delta} + \left(\widehat{Y}^{\delta}\right)^2 {\cal
    C}_A \right]^{-1} \left(\widehat{\bold F}^\delta - {\bold
  F}^\delta \right).
\end{equation}

Finally, we estimated the value of $T^\delta_{\rm CMB}$ by
marginalizing over $F^\delta_{\rm rad}$, considering a flat prior
$-5\%<F^\delta_{\rm rad}<15\%$ (as observed in Sect.~\ref{radconta});
and by computing the first-order moment of the likelihood function,
${\cal L} = e^{-\chi^2/2}$ with respect to $T^\delta_{\rm
  CMB}$\footnote{Note that the obtained $T_{\rm CMB}$ value only
  differs at the third decimal when the maximum likelihood is used as
  estimator.}.

We computed the uncertainties on $T^\delta_{\rm CMB}$ using the second-order moment of ${\cal L}$. Note that the uncertainties only differ at
the fourth decimal at 68\% confidence level. We also computed the
covariance between each $T^\delta_{\rm CMB}$ and 
separated the statistical uncertainties that are fully uncorrelated
between redshift bins, and the systematic errors that are fully correlated
from one bin to another.

Table~\ref{tabdata} summarizes our results for the derived $T_{\rm
  CMB}(z)$ and the associated statistical and systematic
uncertainties.  We verified that removing clusters contaminated by
very bright radio sources ($S \geq 250$~mJy at 1.4~GHz) does not
affect our results.

\section{Results}
\label{sec_res}
The measurement of the tSZ emission law allows us to constrain
the value of $T_{\rm CMB}$ at $z=0$ and its redshift dependance.  We
explore these two constraints separately. We present the results for
the sample of {\it Planck} clusters, which constitutes our base
dataset. We also discuss other cluster samples and finally we give the
tightest constraints obtained by combining tSZ measures from {\it
  Planck} clusters and measures from molecular and atomic absorptions.

\subsection{Constraints on CMB temperature}

\begin{figure*}[!th]
\begin{center}
\includegraphics[width=18cm,trim = 5cm 0cm 0cm 0cm]{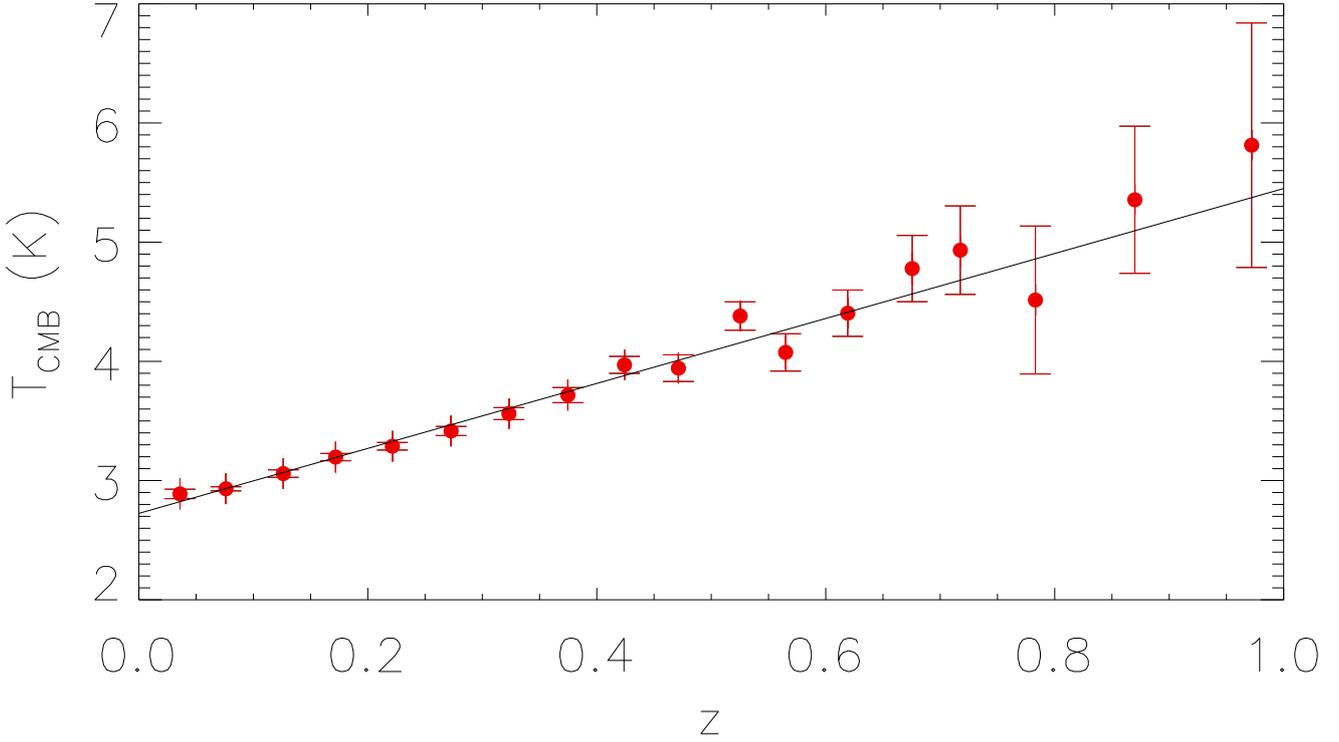}
\caption{Measured $T^\delta_{\rm CMB}$ from the {\it Planck} data are
  shown as red filled circles. The theoretical $T_{\rm CMB}(z)$
  dependance in adiabatic expansion with $T_{\rm CMB}(z=0) = 2.726$~K
  is presented as black solid line. Note that the line is not a fit to the
  data points. Here, error bars only account for the statistical
  dispersion.  The uncertainties due to spectral responses are not
  displayed.}
\label{tcmbz}
\end{center}
\end{figure*}

We computed the $\chi^2$ per degree of freedom (dof) between our
measurements of $T^\delta_{\rm CMB}$ (red filled circles in
Fig.~\ref{tcmbz}) and the adiabatic evolution of the CMB temperature
(solid line in Fig.~\ref{tcmbz}), with $T_{\rm CMB}(z=0) = 2.726 \pm
0.001$~K. We obtained $\chi^2_{dof} = 0.89$. This implies that our
measurements are consistent with the standard $T_{\rm CMB}$ evolution.

\begin{table}
\center
\caption{Measured value of $T^\delta_{\rm CMB}$ in Kelvin derived from
  the tSZ emission law per redshift bin. Systematic errors are fully
  correlated from one redshift bin to the next.}
\label{tabdata}
\begin{tabular}{|c|c|c|c|c|}
\hline
Redshift bin $\delta$ & $N_{\rm cl}$& $T^\delta_{\rm CMB}$ & $\Delta T_{\rm CMB}$ (stat) & $\Delta T_{\rm CMB}$ (syst) \\ \hline
0.00-0.05 & \phantom{0}43 & 2.888 & 0.039 & 0.011\\
0.05-0.10 & 125 & 2.931 & 0.017  & 0.011\\
0.10-0.15 & \phantom{0}92 & 3.059 & 0.032  & 0.012\\
0.15-0.20 & 104 & 3.197 & 0.030  & 0.012\\
0.20-0.25 & \phantom{0}95 & 3.288 & 0.032  & 0.013\\
0.25-0.30 & \phantom{0}87 & 3.416 & 0.038  & 0.013\\
0.30-0.35 & \phantom{0}81 & 3.562 & 0.050  & 0.014\\
0.35-0.40 & \phantom{0}50 & 3.717 & 0.063  & 0.014\\
0.40-0.45 & \phantom{0}45 & 3.971 & 0.071  & 0.015\\
0.45-0.50 & \phantom{0}26 & 3.943 & 0.112  & 0.015\\
0.50-0.55 & \phantom{0}20 & 4.380 & 0.119  & 0.016\\
0.55-0.60 & \phantom{0}18 & 4.075 & 0.156  & 0.016\\
0.60-0.65 & \phantom{0}12 & 4.404 & 0.194  & 0.016\\
0.65-0.70 & \phantom{00}6 & 4.779 & 0.278  & 0.017\\
0.70-0.75 & \phantom{00}5 & 4.933 & 0.371  & 0.017\\
0.75-0.80 & \phantom{00}2 & 4.515 & 0.621  & 0.018\\
0.80-0.85 & \phantom{00}0 &  ---    & --- & --- \\
0.85-0.90 & \phantom{00}1 & 5.356 & 0.617  & 0.019\\
0.90-0.95 & \phantom{00}0 & ---     & --- & --- \\
0.95-1.00 & \phantom{00}1 & 5.813 & 1.025  & 0.020\\ \hline
\end{tabular}
\end{table}

\begin{figure*}[!th]
\begin{center}
\includegraphics[width=18cm,trim = 1cm 0cm 0cm 0cm]{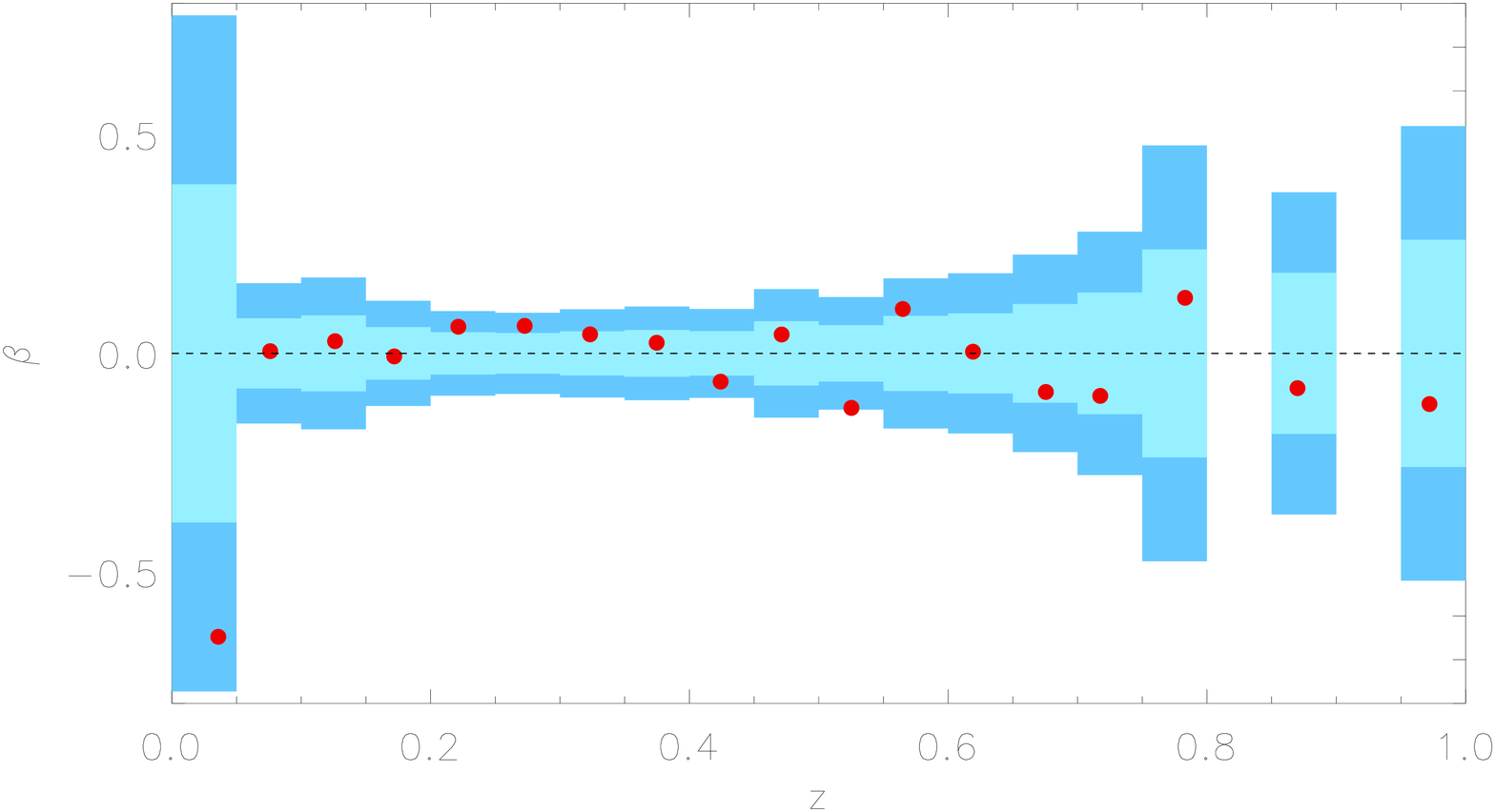}
\caption{Derived $\beta$ values for each redshift bin $\delta$
  considering $T_{\rm CMB}(z=0) = 2.726$. Red filled circles show our
  measurements, light- and dark-blue shaded regions show the 1 and
  2$\sigma$ levels. Errors bars are displayed centered
  on $\beta=0$ to facilitate the comparison between uncertainties in the
  different redshift bins $\delta$. The dashed black line represents $\beta=0$ for standard evolution.}
\label{bz}
\end{center}
\end{figure*}

If we assume an adiabatic expansion, $T_{\rm CMB}(z)$ is written as
\begin{equation}
T_{\rm CMB}(z) = T_{\rm CMB}(z=0)(1+z).
\end{equation} 
We derived $T_{\rm CMB}(z=0)$ from the estimated $T^\delta_{\rm
  CMB}$ for all the redshift bins $\delta$ probed by our sample of {\it
  Planck} clusters. The measurements are presented in Fig.~\ref{tcmbz}
as red filled circles. We observe that all measurements are within a
2$\sigma$ from the adiabatic expansion evolution. The best fit, using
all redshift bins, gives $T_{\rm CMB}(z=0) = 2.720 \pm 0.009 \pm
0.011$~K with statistical and systematics uncertainties,
respectively. Note that the errors are dominated by the systematic
uncertainty from the spectral responses. It is fully correlated
between redshift bins, and thus cannot be reduced. Our measurement of
$T_{\rm CMB}(z=0)$ from the tSZ emission law cannot compete, in terms of
accuracy, with the COBE-FIRAS $T_{\rm CMB}(z=0) = 2.726 \pm 0.001$~K
\citep{fix09}. However, it is fully consistent.

\begin{figure*}[!th]
\begin{center}
\includegraphics[width=18cm,trim = 2.5cm 0cm 0cm 0cm]{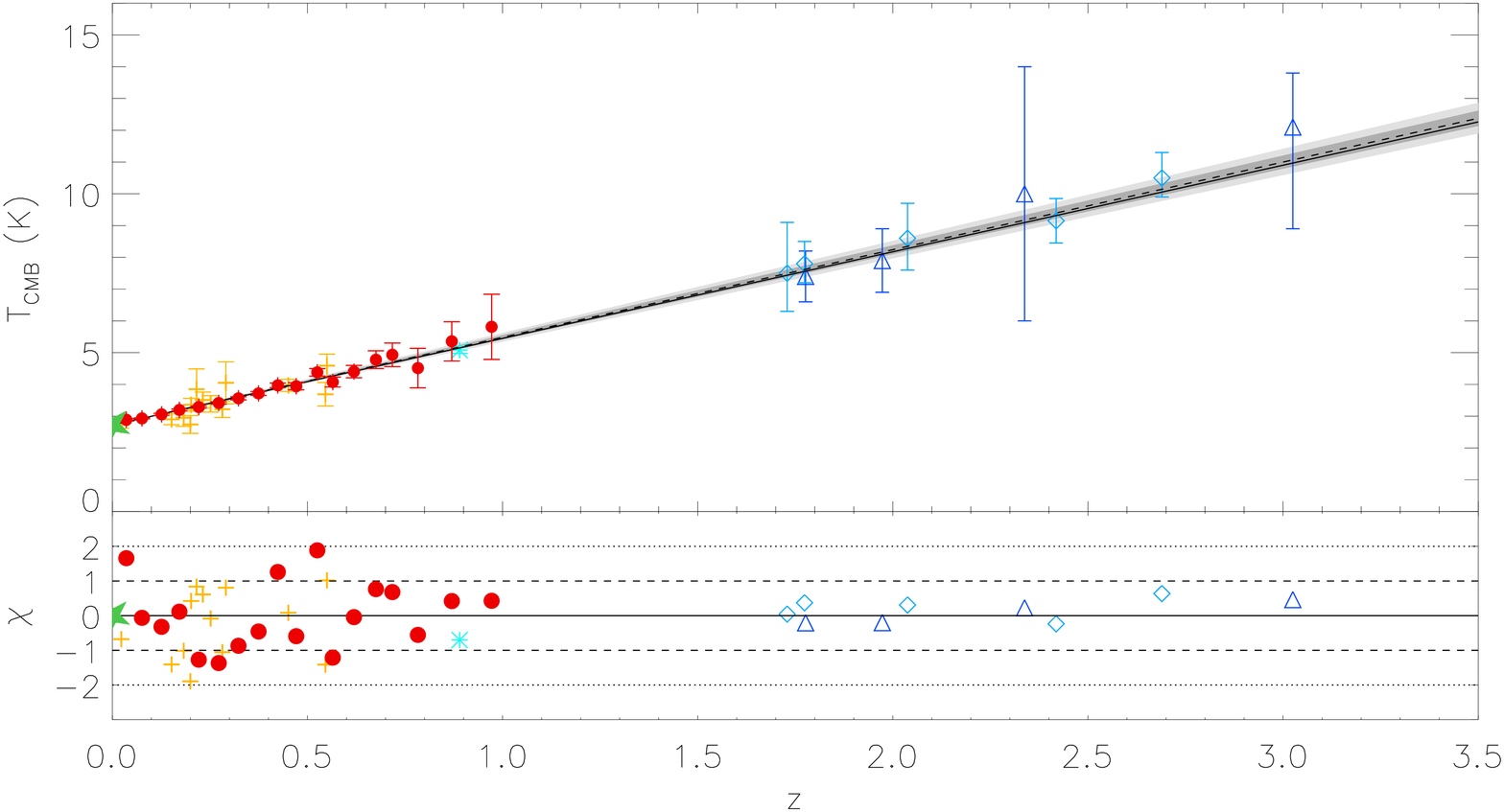}
\caption{{\it Top panel:} $T_{\rm CMB}$ as a function of redshift. The
  red filled circles represent $T_{\rm CMB}$ measured from the tSZ
  emission law in redshift bins of {\it Planck} clusters.  The green
  star shows COBE-FIRAS measurement at $z=0$ \citep{fix09}. The orange
  crosses show $T_{\rm CMB}$ measurements using individual clusters
  \citep{bat02,luz09}.  Dark-blue triangles represent measurements from
  C$_{\rm I}$ and C$_{\rm II}$ absorption
  \citep{cui05,ge97,sri00,mol02} at $z$=(1.8, 2.0, 2.3, 3.0). Blue diamonds show the measurements from CO absorption
  lines \citep{sri08,not11}, and finally the light-blue asterisk is the
  constraint from various molecular species analyses by \citet{mul13}.
  The solid black line presents the standard evolution for $T_{\rm
    CMB}$ and the dashed black line represents our best-fitting model
  combining all the measurements. The 1 and $2\,\sigma$ envelopes are
  displayed as shaded dark and light-gray regions. {\it Bottom
    panel:} Deviation from the standard evolution in units of
  standard deviation. The dashed and dotted black lines correspond to
  the 1 and $2\,\sigma$ levels.}
\label{allexpe}
\end{center}
\end{figure*}

\subsection{Constraints on the redshift evolution}

Our analysis was based on a sample of 813 {\it Planck} clusters
covering a redshift range from 0 to 1. If we express $T_{\rm CMB}(z)$
as in \citet{lim00},
\begin{equation}
T_{\rm CMB}(z) = T_{\rm CMB}(z=0)(1+z)^{1-\beta},
\end{equation} 
it is possible to test the $T_{\rm CMB}$ evolution in adiabatic expansion.
We fit $\beta$ using a maximum-likelihood analysis and a flat prior.
Figure~\ref{bz} presents the best-fitting value of $\beta$ for each
individual redshift bin $\delta$, considering $T_{\rm CMB}(z=0) =
2.726$. We observe that all redshift bins are consistent with $\beta =
0$ at $2\,\sigma$ and that redshift bins from $z=0.05$ to $z=0.7$
provide similar constraints on $\beta$.  Figure~\ref{bz} also
illustrates that our constraint on $\beta$ are not dominated by a
given bin, but that it uses the entire redshift range.

We derived $\beta=0.009\pm0.017$ from the tSZ emission law in the 18
redshift bins used in the analysis, consistent with no deviation from
adiabatic evolution. Note that in contrast to the $T_{\rm CMB}(z=0)$
measurement, $\beta$ is not affected by bandpass uncertainties because
they are fully correlated for all redshift bins.

We compared our result on $\beta$ with the different constraints
obtained either from other tSZ measurements or molecular and atomic
absorptions. \\
Previous analyses have measured
$T_{\rm CMB}(z)$ either at low redshift \citep{luz09} using the tSZ
effect on a small number of clusters, or at higher redshifts
\citep{cui05,ge97,sri00,mol02,mul13} using atomic and molecular
absorption. These $T_{\rm CMB}$ measurements are presented in
Fig.~\ref{allexpe} with the standard adiabatic evolution shown as
a solid black line. \\
We derived for the above-mentioned measurements constraints on $\beta$
using the same fitting procedure and a flat prior as in our
analysis.  Table~\ref{merres} summarizes the results.  We found that
the tSZ emission law from {\it Planck} clusters provides the tightest
constraint on $\beta$. Errors are smaller by about a factor of two
compared with previous constraints, separately.  Our result,
$\beta=0.009\pm0.017$, from tSZ effect in {\it Planck} clusters is at
the same level of precision and accuracy as the tightest constraint,
$\beta=0.009\pm0.019$, reported by \citet{mul13}, combining all the
constraints from molecular and atomic lines and tSZ clusters before
{\it Planck}.\\

By furthermore combining our new limit on $\beta$ with the previous data sets, we improved the
derived constraint on the $T_{\rm CMB}$ evolution. We found $\beta = 0.006
\pm 0.013$. It is worth noting that in this combination the fit is
driven by our own analysis and the measurements from \citet{mul13} and
\citet{not11} which have uncertainties of 0.017, 0.031, and 0.033,
respectively.

\begin{table*}
\center
\caption{Measured values of $T_{\rm CMB}$ and $\beta$ re-derived, with our fitting procedure, from different subsamples of $T_{\rm CMB}(z)$ measurements
  based on the tSZ effect and atomic/molecular absorptions. We provide the $\chi^2_{dof}$ with respect to the standard adiabatic evolution with $T_{\rm CMB} = 2.726 \pm 0.001$~K and $\beta=0$. }
\label{merres}
\begin{tabular}{|p{3.5cm}|>{\centering}p{1cm}|>{\centering}p{1cm}|>{\centering}p{1cm}|>{\centering}p{1cm}|>{\centering}p{1cm}|>{\centering}p{1cm}|>{\centering}p{1.5cm}|}
\hline
Measurement method & $T_{\rm CMB}$ & $\Delta T_{\rm CMB}$ & $\beta$ & $\Delta \beta$ & $\chi^2_{dof}$ & dof & References\tabularnewline
\hline
${\rm C}_{\rm I}$ and ${\rm C}_{\rm II}$ absorption & 2.71 & 0.20 & -0.002 & 0.067 & 0.09 & \leavevmode\hphantom{0}4 & (a)\tabularnewline
CO absorption & 2.78 & 0.11 & -0.018 & 0.033 & 0.14 & \leavevmode\hphantom{0}5 & (b) \tabularnewline
Various molecular species & 2.69 & 0.05 & 0.022 & 0.031 & 0.52 & \leavevmode\hphantom{0}1 & (c) \tabularnewline
tSZ  & 2.66 & 0.05 & 0.034 & 0.078 & 1.01 & 13 & (d) \tabularnewline
tSZ (this work) & 2.72 & 0.01 & 0.009 & 0.017 & 0.89 & 18 &(e) \tabularnewline
\hline
All atomic and molecular & 2.71 & 0.05 & 0.003 & 0.021 & 0.16 & 10 & (a,b,c) \tabularnewline
All tSZ & 2.72 & 0.01 & 0.009 & 0.016& 0.91 & 31 & (d,e) \tabularnewline
\hline
All  & 2.72 & 0.01 & 0.006 & 0.013 &0.74 & 41 & (a,b,c,d,e) \tabularnewline
\hline
\end{tabular}\\
\begin{tabular}{l}
(a) using data samples from \citet{cui05,ge97,sri00} and \citet{mol02}.\\
(b) using data samples from \citet{sri08} and \citet{not11}.\\
(c) using data samples from \citet{mul13}.\\
(d) using data samples from \citet{bat02} and \citet{luz09}.\\
(e) The present analysis.
\end{tabular}
\end{table*}

\section{Discussion}
\label{sec_disc}

Our basic results presented in Sect.~\ref{sec_res} were derived from the
analysis of tSZ emission law of 813 clusters from {\it Planck} up to
$z\sim1$. We found $\beta=0.009\pm0.017$, which shows that even with a
moderately deep sample tSZ is a competitive observational way of
constraining the evolution of $T_{\rm CMB}$.

To explore the variation of this result with other cluster
samples, we performed the same analysis, as in
Sect.~\ref{sec_ana}, on the MCXC \citep{pif11} catalog. This allowed us to
select a large number of massive clusters (see Table~\ref{tab_cat}). To avoid
significant radio contamination of the tSZ measurement, we considered only clusters with an intensity lower than $500~\mu$K$_{\rm CMB}$ in
the 100~GHz channel. This process only removes three clusters from our
initial MCXC sample (see Sect.~\ref{sec_cat}). We thus performed our analysis, and with the obtained
$T_{\rm CMB}$ measures ranging from $z=0$ to 0.35, we found $\beta =
0.07 \pm 0.10$. This poor constraint is mainly due to the reduced
redshift range and to the relatively low tSZ flux associated with a
large portion of the considered clusters.
We also analyzed the SPT \citep{rei13} and ACT
\citep{has13} catalogs. Again, we derived poor constraints on $\beta$, because the tSZ signal in the {\it Planck} intensity maps from the selected clusters is faint. Consequently, we did not include these measurements in the analysis. Combining the constraints from the tSZ emission law with other direct measurement using molecular and atomic absorption, we found the tightest limit on the $T_{\rm CMB}$ evolution to be $\beta = 0.006
\pm 0.013$, consistent with a standard $T_{\rm CMB}$ evolution.\\

Departures from an adiabatic evolution of $T_{\rm CMB}$ can be tested through indirect measurements. In particular when CMB photon number is not conserved, $\beta$ can be constrained by the distance duality relation violation  \citep[][and references therein]{avg12} or by combining the CMB
and galaxy distribution \citep[as in, e.g., ][]{oph05}. These indirect measurements yield $\beta = 0.010 \pm 0.020$ and $\beta \leq
0.0034$, consistent with our analysis. Our tSZ-based measurements can also test deviations from the $T_{\rm CMB}$-redshift relation in decaying dark-energy (DE) models. \citet{jer11} predicted the following relation:
\begin{eqnarray}
T_{\rm CMB}(z) &=& T_{\rm CMB}(z=0) \times (1+z) \\ \nonumber
	&&\times \left( \frac{(m-3 \Omega_{{\rm m}}) + m(1+z)^{m-3}\Omega_\Lambda}{(m-3)\Omega_{\rm m}} \right)^{1/3},
\end{eqnarray}
with $\Omega_{\rm m}$ and $\Omega_\Lambda$ the matter and DE energy densities at $z=0$, and $m=3(w_{\rm eff}+1)$ is related to the effective equation of state of the decaying DE, $w_{\rm eff}$.
Considering a flat Universe and $\Omega_{\rm m} = 0.314 \pm 0.020$ \citep{PlanckPAR}, we derive $w_{\rm eff}=-0.995\pm0.011$, which improves previous
constraint from \citet[e.g., ][]{not11}.\\

The evolution of $T_{\rm CMB}$ with redshift can also be affected by time-varying fundamental constants \citep[see][for a recent review]{uza11}.
Specifically for the tSZ-based constraints, variations of $h$ and $k_{\rm B}$ in Eq.~\ref{szspec}  lead to deviations of $\beta$ from zero $\frac{h}{k_{\rm B} T_{\rm CMB}(0)}(1+z)^{\beta} = {\rm C^{ste}}$. 
However, variations of the fundamental constants become small after the Universe enters its current DE-dominated epoch \citep{bar02}, and consequently tSZ variations are less sensitive to these changes at $z < 1 $.\\

\section{Conclusion}
\label{sec_concl}

We have performed an analysis of the {\it Planck} intensity data in the
range of 100 to 857~GHz, aimed at deriving the CMB temperature and its evolution
via the tSZ emission law. Based on the {\it Planck} SZ catalog, we measured $T_{\rm CMB}(z)$
in the redshift range $0<z<1$. This is the first
measurement of $T_{\rm CMB}(z)$ on such a large tSZ sample of clusters. 
It demonstrates the ability of exploring the low-redshift range which cannot be covered
by the traditionally used optical/UV quasar absorption systems.\\

We showed that clustered CIB and relativistic corrections to the
tSZ spectral law do not produce any significant bias on our result. We note that
the main uncertainties are caused by the {\it Planck} instrumental
 spectral responses, CMB contamination, and the radio source contamination. 
 They were modeled and accounted for in the error bars in the determination of $T_{\rm CMB}(z)$.\\

Our measurement of $T_{\rm CMB}(z)$ below $z=1$ reaches a precision of
about 5\% below $z=0.65$, about 1\% for all bins below $z=0.3$, and better
than 0.6\% for the redshift between $z=0.05$ and $z=0.10$. This is
the most precise measurement at $z>0$ to date.
Combining $T_{\rm CMB}(z)$ at low redshifts with results from \citet{mul13} and
  references presented in Fig.~\ref{allexpe}, we obtained the
tightest constraints so far on the $T_{\rm CMB}(z) = T_{\rm
  CMB}(z=0)(1+z)^{1-\beta}$ law, with $\beta = 0.006 \pm 0.013$.
Our result confirms that the CMB temperature evolution is
consistent with an adiabatic
expansion.\\

The discovery of new massive high-$z$ ($z>0.5$) clusters will bring a major improvement in $T_{\rm CMB}(z)$ 
measurement from the tSZ emission law.  In particular, SZ surveys such as Planck and SPT-3G and optical surveys such as
the Dark Energy Survey \citep[][]{des05} and 
Pan-STARRS \citep{kai02,ton12} will provide us with much larger and deeper cluster samples. In the future,  even 
larger and deeper samples of massive clusters will be constructed from from  \textit{EUCLID} \citep{ami12,ame12}, the Large Synoptic Survey
Telescope \citep[][]{lsst}, and SRG-eROSITA \citep{mer12,pil12}. In combination with the fourth-generation CMB space mission, the tSZ emission
law from clusters will strongly improve existing constraints on $T_{\rm CMB}(z)$ up to $z=2$.

\section*{Acknowledgements}
\thanks{
    We thank the anonymous referee for his or her comments. We are grateful to J.F. Mac\'ias-P\'erez, J.M. Diego, and R.Genova-Santos for their comments and suggestions.
    Some of the results in this paper have been derived using the {\tt HEALPix}
  package \citep{gor05}.\\
  We acknowledge the support of the French \emph{Agence
    Nationale de la Recherche} under grant ANR-11-BD56-015.\\ 
    The development of Planck has been supported by: ESA; CNES and CNRS/INSU-IN2P3-INP (France); ASI, CNR, and INAF (Italy); NASA and DoE (USA); STFC and UKSA (UK); CSIC, MICINN and JA (Spain); Tekes, AoF and CSC (Finland); DLR and MPG (Germany); CSA (Canada); DTU Space (Denmark); SER/SSO (Switzerland); RCN (Norway); SFI (Ireland); FCT/MCTES (Portugal); and The development of Planck has been supported by: ESA; CNES and CNRS/INSU-IN2P3-INP (France); ASI, CNR, and INAF (Italy); NASA and DoE (USA); STFC and UKSA (UK); CSIC, MICINN and JA (Spain); Tekes, AoF and CSC (Finland); DLR and MPG (Germany); CSA (Canada); DTU Space (Denmark); SER/SSO (Switzerland); RCN (Norway); SFI (Ireland); FCT/MCTES (Portugal); and PRACE (EU).}

\bibliographystyle{aa}
\bibliography{TCMB_publi.bib}

\begin{thebibliography}{91}
\expandafter\ifx\csname natexlab\endcsname\relax\def\natexlab#1{#1}\fi

\bibitem[{{Addison} {et~al.}(2012){Addison}, {Dunkley}, \& {Spergel}}]{add12}
{Addison}, G.~E., {Dunkley}, J., \& {Spergel}, D.~N. 2012, \mnras, 427, 1741

\bibitem[{{Amendola} {et~al.}(2012){Amendola}, {Appleby}, {Bacon}, {Baker},
  {Baldi}, {Bartolo}, {Blanchard}, {Bonvin}, {Borgani}, {Branchini}, {Burrage},
  {Camera}, {Carbone}, {Casarini}, {Cropper}, {deRham}, {di Porto}, {Ealet},
  {Ferreira}, {Finelli}, {Garcia-Bellido}, {Giannantonio}, {Guzzo}, {Heavens},
  {Heisenberg}, {Heymans}, {Hoekstra}, {Hollenstein}, {Holmes}, {Horst},
  {Jahnke}, {Kitching}, {Koivisto}, {Kunz}, {La Vacca}, {March}, {Majerotto},
  {Markovic}, {Marsh}, {Marulli}, {Massey}, {Mellier}, {Mota}, {Nunes},
  {Percival}, {Pettorino}, {Porciani}, {Quercellini}, {Read}, {Rinaldi},
  {Sapone}, {Scaramella}, {Skordis}, {Simpson}, {Taylor}, {Thomas}, {Trotta},
  {Verde}, {Vernizzi}, {Vollmer}, {Wang}, {Weller}, \& {Zlosnik}}]{ame12}
{Amendola}, L., {Appleby}, S., {Bacon}, D., {et~al.} 2012, ArXiv e-prints

\bibitem[{{Amiaux} {et~al.}(2012){Amiaux}, {Scaramella}, {Mellier}, {Altieri},
  {Burigana}, {Da Silva}, {Gomez}, {Hoar}, {Laureijs}, {Maiorano},
  {Magalh{\~a}es Oliveira}, {Renk}, {Saavedra Criado}, {Tereno},
  {Augu{\`e}res}, {Brinchmann}, {Cropper}, {Duvet}, {Ealet}, {Franzetti},
  {Garilli}, {Gondoin}, {Guzzo}, {Hoekstra}, {Holmes}, {Jahnke}, {Kitching},
  {Meneghetti}, {Percival}, \& {Warren}}]{ami12}
{Amiaux}, J., {Scaramella}, R., {Mellier}, Y., {et~al.} 2012, in Society of
  Photo-Optical Instrumentation Engineers (SPIE) Conference Series, Vol. 8442,
  Society of Photo-Optical Instrumentation Engineers (SPIE) Conference Series

\bibitem[{{Arnaud} {et~al.}(2010){Arnaud}, {Pratt}, {Piffaretti},
  {B{\"o}hringer}, {Croston}, \& {Pointecouteau}}]{arn10}
{Arnaud}, M., {Pratt}, G.~W., {Piffaretti}, R., {et~al.} 2010, \aap, 517, A92

\bibitem[{{Avgoustidis} {et~al.}(2012){Avgoustidis}, {Luzzi}, {Martins}, \&
  {Monteiro}}]{avg12}
{Avgoustidis}, A., {Luzzi}, G., {Martins}, C.~J.~A.~P., \& {Monteiro},
  A.~M.~R.~V.~L. 2012, \jcap, 2, 13

\bibitem[{{Bahcall} \& {Wolf}(1968)}]{bah68}
{Bahcall}, J.~N. \& {Wolf}, R.~A. 1968, \apj, 152, 701

\bibitem[{{Barrow} {et~al.}(2002){Barrow}, {Sandvik}, \& {Magueijo}}]{bar02}
{Barrow}, J.~D., {Sandvik}, H.~B., \& {Magueijo}, J. 2002, \prd, 65, 063504

\bibitem[{{Battistelli} {et~al.}(2002){Battistelli}, {De Petris}, {Lamagna},
  {Melchiorri}, {Palladino}, {Savini}, {Cooray}, {Melchiorri}, {Rephaeli}, \&
  {Shimon}}]{bat02}
{Battistelli}, E.~S., {De Petris}, M., {Lamagna}, L., {et~al.} 2002, \apjl,
  580, L101

\bibitem[{{Bennett} {et~al.}(2003){Bennett}, {Halpern}, {Hinshaw}, {Jarosik},
  {Kogut}, {Limon}, {Meyer}, {Page}, {Spergel}, {Tucker}, {Wollack}, {Wright},
  {Barnes}, {Greason}, {Hill}, {Komatsu}, {Nolta}, {Odegard}, {Peiris},
  {Verde}, \& {Weiland}}]{ben03}
{Bennett}, C.~L., {Halpern}, M., {Hinshaw}, G., {et~al.} 2003, \apjs, 148, 1

\bibitem[{{Chang} {et~al.}(2009){Chang}, {Ade}, {Aird}, {Benson}, {Bleem},
  {Carlstrom}, {Cho}, {de Haan}, {Crawford}, {Crites}, {Dobbs}, {Everett},
  {Halverson}, {Holder}, {Holzapfel}, {Hrubes}, {Joy}, {Keisler}, {Lanting},
  {Lee}, {Leitch}, {Loehr}, {Lueker}, {McMahon}, {Mehl}, {Meyer}, {Mohr},
  {Montroy}, {Ngeow}, {Padin}, {Plagge}, {Pryke}, {Reichardt}, {Ruhl},
  {Schaffer}, {Shaw}, {Shirokoff}, {Spieler}, {Stalder}, {Stark},
  {Staniszewski}, {Vanderlinde}, {Vieira}, {Williamson}, {Zahn}, \&
  {Zenteno}}]{cha09}
{Chang}, C.~L., {Ade}, P.~A.~R., {Aird}, K.~A., {et~al.} 2009, in American
  Institute of Physics Conference Series, Vol. 1185, American Institute of
  Physics Conference Series, ed. B.~{Young}, B.~{Cabrera}, \& A.~{Miller},
  475--477

\bibitem[{{Clarkson}(2012)}]{cla12}
{Clarkson}, C. 2012, Comptes Rendus Physique, 13, 682

\bibitem[{{Combes} \& {Wiklind}(1999)}]{com99}
{Combes}, F. \& {Wiklind}, T. 1999, in Astronomical Society of the Pacific
  Conference Series, Vol. 156, Highly Redshifted Radio Lines, ed. C.~L.
  {Carilli}, S.~J.~E. {Radford}, K.~M. {Menten}, \& G.~I. {Langston}, 210

\bibitem[{{Condon} {et~al.}(1998){Condon}, {Cotton}, {Greisen}, {Yin},
  {Perley}, {Taylor}, \& {Broderick}}]{con98}
{Condon}, J.~J., {Cotton}, W.~D., {Greisen}, E.~W., {et~al.} 1998, \aj, 115,
  1693

\bibitem[{{Cui} {et~al.}(2005){Cui}, {Bechtold}, {Ge}, \& {Meyer}}]{cui05}
{Cui}, J., {Bechtold}, J., {Ge}, J., \& {Meyer}, D.~M. 2005, \apj, 633, 649

\bibitem[{{de Martino} {et~al.}(2012){de Martino}, {Atrio-Barandela}, {da
  Silva}, {Ebeling}, {Kashlinsky}, {Kocevski}, \& {Martins}}]{mar12}
{de Martino}, I., {Atrio-Barandela}, F., {da Silva}, A., {et~al.} 2012, \apj,
  757, 144

\bibitem[{{Fabbri} {et~al.}(1978){Fabbri}, {Melchiorri}, \& {Natale}}]{fab78}
{Fabbri}, R., {Melchiorri}, F., \& {Natale}, V. 1978, \apss, 59, 223

\bibitem[{{Fixsen}(2009)}]{fix09}
{Fixsen}, D.~J. 2009, \apj, 707, 916

\bibitem[{{Fixsen} {et~al.}(1996){Fixsen}, {Cheng}, {Gales}, {Mather},
  {Shafer}, \& {Wright}}]{fix96}
{Fixsen}, D.~J., {Cheng}, E.~S., {Gales}, J.~M., {et~al.} 1996, \apj, 473, 576

\bibitem[{{Freese} {et~al.}(1987){Freese}, {Adams}, {Frieman}, \&
  {Mottola}}]{fre87}
{Freese}, K., {Adams}, F.~C., {Frieman}, J.~A., \& {Mottola}, E. 1987, Nuclear
  Physics B, 287, 797

\bibitem[{{Ge} {et~al.}(1997){Ge}, {Bechtold}, \& {Black}}]{ge97}
{Ge}, J., {Bechtold}, J., \& {Black}, J.~H. 1997, \apj, 474, 67

\bibitem[{{Goodman}(1995)}]{goo95}
{Goodman}, J. 1995, \prd, 52, 1821

\bibitem[{{G{\'o}rski} {et~al.}(2005){G{\'o}rski}, {Hivon}, {Banday},
  {Wandelt}, {Hansen}, {Reinecke}, \& {Bartelmann}}]{gor05}
{G{\'o}rski}, K.~M., {Hivon}, E., {Banday}, A.~J., {et~al.} 2005, \apj, 622,
  759

\bibitem[{{Hasselfield} {et~al.}(2013){Hasselfield}, {Hilton}, {Marriage},
  {Addison}, {Barrientos}, {Battaglia}, {Battistelli}, {Bond}, {Crichton},
  {Das}, {Devlin}, {Dicker}, {Dunkley}, {Dunner}, {Fowler}, {Gralla}, {Hajian},
  {Halpern}, {Hincks}, {Hlozek}, {Hughes}, {Infante}, {Irwin}, {Kosowsky},
  {Marsden}, {Menanteau}, {Moodley}, {Niemack}, {Nolta}, {Page}, {Partridge},
  {Reese}, {Schmitt}, {Sehgal}, {Sherwin}, {Sievers}, {Sif{\'o}n}, {Spergel},
  {Staggs}, {Swetz}, {Switzer}, {Thornton}, {Trac}, \& {Wollack}}]{has13}
{Hasselfield}, M., {Hilton}, M., {Marriage}, T.~A., {et~al.} 2013, e-prints
  ArXiv:1301.0816

\bibitem[{{Hauser} \& {Dwek}(2001)}]{hau01}
{Hauser}, M.~G. \& {Dwek}, E. 2001, \araa, 39, 249

\bibitem[{{Hincks} {et~al.}(2010){Hincks}, {Acquaviva}, {Ade}, {Aguirre},
  {Amiri}, {Appel}, {Barrientos}, {Battistelli}, {Bond}, {Brown}, {Burger},
  {Chervenak}, {Das}, {Devlin}, {Dicker}, {Doriese}, {Dunkley}, {D{\"u}nner},
  {Essinger-Hileman}, {Fisher}, {Fowler}, {Hajian}, {Halpern}, {Hasselfield},
  {Hern{\'a}ndez-Monteagudo}, {Hilton}, {Hilton}, {Hlozek}, {Huffenberger},
  {Hughes}, {Hughes}, {Infante}, {Irwin}, {Jimenez}, {Juin}, {Kaul}, {Klein},
  {Kosowsky}, {Lau}, {Limon}, {Lin}, {Lupton}, {Marriage}, {Marsden},
  {Martocci}, {Mauskopf}, {Menanteau}, {Moodley}, {Moseley}, {Netterfield},
  {Niemack}, {Nolta}, {Page}, {Parker}, {Partridge}, {Quintana}, {Reid},
  {Sehgal}, {Sievers}, {Spergel}, {Staggs}, {Stryzak}, {Swetz}, {Switzer},
  {Thornton}, {Trac}, {Tucker}, {Verde}, {Warne}, {Wilson}, {Wollack}, \&
  {Zhao}}]{hin10}
{Hincks}, A.~D., {Acquaviva}, V., {Ade}, P.~A.~R., {et~al.} 2010, \apjs, 191,
  423

\bibitem[{{Horellou} {et~al.}(2005){Horellou}, {Nord}, {Johansson}, \&
  {L{\'e}vy}}]{hor05}
{Horellou}, C., {Nord}, M., {Johansson}, D., \& {L{\'e}vy}, A. 2005, \aap, 441,
  435

\bibitem[{{Hurier} {et~al.}(2013){Hurier}, {Hildebrandt}, \&
  {Macias-Perez}}]{hur13}
{Hurier}, G., {Hildebrandt}, S.~R., \& {Macias-Perez}, J.~F. 2013, e-prints
  ArXiv:1007.1149

\bibitem[{{Itoh} {et~al.}(1998){Itoh}, {Kohyama}, \& {Nozawa}}]{ito98}
{Itoh}, N., {Kohyama}, Y., \& {Nozawa}, S. 1998, \apj, 502, 7

\bibitem[{{Jaeckel} \& {Ringwald}(2010)}]{jae10}
{Jaeckel}, J. \& {Ringwald}, A. 2010, Annual Review of Nuclear and Particle
  Science, 60, 405

\bibitem[{{Jetzer} {et~al.}(2011){Jetzer}, {Puy}, {Signore}, \&
  {Tortora}}]{jer11}
{Jetzer}, P., {Puy}, D., {Signore}, M., \& {Tortora}, C. 2011, General
  Relativity and Gravitation, 43, 1083

\bibitem[{{Kaiser} {et~al.}(2002){Kaiser}, {Aussel}, {Burke}, {Boesgaard},
  {Chambers}, {Chun}, {Heasley}, {Hodapp}, {Hunt}, {Jedicke}, {Jewitt},
  {Kudritzki}, {Luppino}, {Maberry}, {Magnier}, {Monet}, {Onaka}, {Pickles},
  {Rhoads}, {Simon}, {Szalay}, {Szapudi}, {Tholen}, {Tonry}, {Waterson}, \&
  {Wick}}]{kai02}
{Kaiser}, N., {Aussel}, H., {Burke}, B.~E., {et~al.} 2002, in Society of
  Photo-Optical Instrumentation Engineers (SPIE) Conference Series, Vol. 4836,
  Society of Photo-Optical Instrumentation Engineers (SPIE) Conference Series,
  ed. J.~A. {Tyson} \& S.~{Wolff}, 154--164

\bibitem[{{Kashlinsky}(2005)}]{kas05}
{Kashlinsky}, A. 2005, \physrep, 409, 361

\bibitem[{{Lagache} {et~al.}(2005){Lagache}, {Puget}, \& {Dole}}]{lag05}
{Lagache}, G., {Puget}, J.-L., \& {Dole}, H. 2005, \araa, 43, 727

\bibitem[{{Lima} {et~al.}(2000){Lima}, {Silva}, \& {Viegas}}]{lim00}
{Lima}, J.~A.~S., {Silva}, A.~I., \& {Viegas}, S.~M. 2000, \mnras, 312, 747

\bibitem[{{Lima} \& {Trodden}(1996)}]{lim96}
{Lima}, J.~A.~S. \& {Trodden}, M. 1996, \prd, 53, 4280

\bibitem[{{Losecco} {et~al.}(2001){Losecco}, {Mathews}, \& {Wang}}]{los01}
{Losecco}, J.~M., {Mathews}, G.~J., \& {Wang}, Y. 2001, \prd, 64, 123002

\bibitem[{{LSST Science Collaboration} {et~al.}(2009){LSST Science
  Collaboration}, {Abell}, {Allison}, {Anderson}, {Andrew}, {Angel}, {Armus},
  {Arnett}, {Asztalos}, {Axelrod}, \& et~al.}]{lsst}
{LSST Science Collaboration}, {Abell}, P.~A., {Allison}, J., {et~al.} 2009,
  ArXiv e-prints

\bibitem[{{Lu} {et~al.}(1996){Lu}, {Sargent}, {Barlow}, {Churchill}, \&
  {Vogt}}]{lu96}
{Lu}, L., {Sargent}, W.~L.~W., {Barlow}, T.~A., {Churchill}, C.~W., \& {Vogt},
  S.~S. 1996, \apjs, 107, 475

\bibitem[{{Luzzi} {et~al.}(2009){Luzzi}, {Shimon}, {Lamagna}, {Rephaeli}, {De
  Petris}, {Conte}, {De Gregori}, \& {Battistelli}}]{luz09}
{Luzzi}, G., {Shimon}, M., {Lamagna}, L., {et~al.} 2009, \apj, 705, 1122

\bibitem[{{Mather} {et~al.}(1999){Mather}, {Fixsen}, {Shafer}, {Mosier}, \&
  {Wilkinson}}]{mat99}
{Mather}, J.~C., {Fixsen}, D.~J., {Shafer}, R.~A., {Mosier}, C., \&
  {Wilkinson}, D.~T. 1999, \apj, 512, 511

\bibitem[{{Matyjasek}(1995)}]{maty95}
{Matyjasek}, J. 1995, \prd, 51, 4154

\bibitem[{{Mauch} {et~al.}(2003){Mauch}, {Murphy}, {Buttery}, {Curran},
  {Hunstead}, {Piestrzynski}, {Robertson}, \& {Sadler}}]{mau03}
{Mauch}, T., {Murphy}, T., {Buttery}, H.~J., {et~al.} 2003, \mnras, 342, 1117

\bibitem[{{Mauch} {et~al.}(2008){Mauch}, {Murphy}, {Buttery}, {Curran},
  {Hunstead}, {Piestrzynski}, {Ropbertson}, \& {Sadler}}]{mau08}
{Mauch}, T., {Murphy}, T., {Buttery}, H.~J., {et~al.} 2008, VizieR Online Data
  Catalog, 8081, 0

\bibitem[{{Merloni} {et~al.}(2012){Merloni}, {Predehl}, {Becker},
  {B{\"o}hringer}, {Boller}, {Brunner}, {Brusa}, {Dennerl}, {Freyberg},
  {Friedrich}, {Georgakakis}, {Haberl}, {Hasinger}, {Meidinger}, {Mohr},
  {Nandra}, {Rau}, {Reiprich}, {Robrade}, {Salvato}, {Santangelo}, {Sasaki},
  {Schwope}, {Wilms}, \& {German eROSITA Consortium}}]{mer12}
{Merloni}, A., {Predehl}, P., {Becker}, W., {et~al.} 2012, ArXiv e-prints

\bibitem[{{Meyer} {et~al.}(1986){Meyer}, {York}, {Black}, {Chaffee}, \&
  {Foltz}}]{mey86}
{Meyer}, D.~M., {York}, D.~G., {Black}, J.~H., {Chaffee}, Jr., F.~H., \&
  {Foltz}, C.~B. 1986, \apjl, 308, L37

\bibitem[{{Molaro} {et~al.}(2002){Molaro}, {Levshakov}, {Dessauges-Zavadsky},
  \& {D'Odorico}}]{mol02}
{Molaro}, P., {Levshakov}, S.~A., {Dessauges-Zavadsky}, M., \& {D'Odorico}, S.
  2002, \aap, 381, L64

\bibitem[{{Muller} {et~al.}(2013){Muller}, {Beelen}, {Black}, {Curran},
  {Horellou}, {Aalto}, {Combes}, {Gu{\'e}lin}, \& {Henkel}}]{mul13}
{Muller}, S., {Beelen}, A., {Black}, J.~H., {et~al.} 2013, \aap, 551, A109

\bibitem[{{Murphy} {et~al.}(2003){Murphy}, {Webb}, \& {Flambaum}}]{mur03}
{Murphy}, M.~T., {Webb}, J.~K., \& {Flambaum}, V.~V. 2003, \mnras, 345, 609

\bibitem[{{Noterdaeme} {et~al.}(2011){Noterdaeme}, {Petitjean}, {Srianand},
  {Ledoux}, \& {L{\'o}pez}}]{not11}
{Noterdaeme}, P., {Petitjean}, P., {Srianand}, R., {Ledoux}, C., \&
  {L{\'o}pez}, S. 2011, \aap, 526, L7

\bibitem[{{Nozawa} {et~al.}(2000){Nozawa}, {Itoh}, {Kawana}, \&
  {Kohyama}}]{ito00}
{Nozawa}, S., {Itoh}, N., {Kawana}, Y., \& {Kohyama}, Y. 2000, \apj, 536, 31

\bibitem[{{Opher} \& {Pelinson}(2005)}]{oph05}
{Opher}, R. \& {Pelinson}, A. 2005, Brazilian Journal of Physics, 35, 1206

\bibitem[{{Overduin} \& {Cooperstock}(1998)}]{ove98}
{Overduin}, J.~M. \& {Cooperstock}, F.~I. 1998, \prd, 58, 043506

\bibitem[{{Piffaretti} {et~al.}(2011){Piffaretti}, {Arnaud}, {Pratt},
  {Pointecouteau}, \& {Melin}}]{pif11}
{Piffaretti}, R., {Arnaud}, M., {Pratt}, G.~W., {Pointecouteau}, E., \&
  {Melin}, J.-B. 2011, \aap, 534, A109

\bibitem[{{Pillepich} {et~al.}(2012){Pillepich}, {Porciani}, \&
  {Reiprich}}]{pil12}
{Pillepich}, A., {Porciani}, C., \& {Reiprich}, T.~H. 2012, \mnras, 422, 44

\bibitem[{{Planck Collaboration early VIII}(2011)}]{PlanckESZ}
{Planck Collaboration early VIII}. 2011, \aap, 536, A8

\bibitem[{{Planck Collaboration early XIX}(2011)}]{PlanckDUST}
{Planck Collaboration early XIX}. 2011, \aap, 536, A19

\bibitem[{{Planck Collaboration early XVIII}(2011)}]{PlanckCIB}
{Planck Collaboration early XVIII}. 2011, \aap, 536, A18

\bibitem[{{Planck Collaboration early XXVI}(2011)}]{PlanckHZ}
{Planck Collaboration early XXVI}. 2011, \aap, 536, A26

\bibitem[{{Planck Collaboration int. V}(2013)}]{PlanckPPP}
{Planck Collaboration int. V}. 2013, \aap, 550, A131

\bibitem[{{Planck Collaboration int. XIII}(2013)}]{PlanckkSZ}
{Planck Collaboration int. XIII}. 2013, e-prints ArXiv:1303.5090

\bibitem[{{Planck Collaboration int. XIV}(2013)}]{PlanckDUST2}
{Planck Collaboration int. XIV}. 2013, e-prints ArXiv:1307.6815

\bibitem[{{Planck Collaboration results I}(2013)}]{PlanckMIS}
{Planck Collaboration results I}. 2013, e-prints ArXiv:1303.5062

\bibitem[{{Planck Collaboration results IX}(2013)}]{PlanckBP}
{Planck Collaboration results IX}. 2013, e-prints ArXiv:1303.5070

\bibitem[{{Planck Collaboration results VI}(2013)}]{PlanckDPC}
{Planck Collaboration results VI}. 2013, e-prints ArXiv:1303.5067

\bibitem[{{Planck Collaboration results VII}(2013)}]{PlanckBEAM}
{Planck Collaboration results VII}. 2013, e-prints ArXiv:1303.5068

\bibitem[{{Planck Collaboration results VIII}(2013)}]{PlanckCAL}
{Planck Collaboration results VIII}. 2013, e-prints ArXiv:1303.5069

\bibitem[{{Planck Collaboration results XV}(2013)}]{PlanckPS}
{Planck Collaboration results XV}. 2013, e-prints ArXiv:1303.5075

\bibitem[{{Planck Collaboration results XVI}(2013)}]{PlanckPAR}
{Planck Collaboration results XVI}. 2013, e-prints ArXiv:1303.5076

\bibitem[{{Planck Collaboration results XXI}(2013)}]{PlanckSZS}
{Planck Collaboration results XXI}. 2013, e-prints ArXiv:1303.5081

\bibitem[{{Planck Collaboration results XXIX}(2013)}]{PlanckSZC}
{Planck Collaboration results XXIX}. 2013, e-prints ArXiv:1303.5089

\bibitem[{{Planck Collaboration results XXVIII}(2013)}]{PlanckCCS}
{Planck Collaboration results XXVIII}. 2013, e-prints ArXiv:1303.5088

\bibitem[{{Pointecouteau} {et~al.}(1998){Pointecouteau}, {Giard}, \&
  {Barret}}]{poi98}
{Pointecouteau}, E., {Giard}, M., \& {Barret}, D. 1998, \aap, 336, 44

\bibitem[{{Puy}(2004)}]{puy04}
{Puy}, D. 2004, \aap, 422, 1

\bibitem[{{Reichardt} {et~al.}(2013){Reichardt}, {Stalder}, {Bleem}, {Montroy},
  {Aird}, {Andersson}, {Armstrong}, {Ashby}, {Bautz}, {Bayliss}, {Bazin},
  {Benson}, {Brodwin}, {Carlstrom}, {Chang}, {Cho}, {Clocchiatti}, {Crawford},
  {Crites}, {de Haan}, {Desai}, {Dobbs}, {Dudley}, {Foley}, {Forman}, {George},
  {Gladders}, {Gonzalez}, {Halverson}, {Harrington}, {High}, {Holder},
  {Holzapfel}, {Hoover}, {Hrubes}, {Jones}, {Joy}, {Keisler}, {Knox}, {Lee},
  {Leitch}, {Liu}, {Lueker}, {Luong-Van}, {Mantz}, {Marrone}, {McDonald},
  {McMahon}, {Mehl}, {Meyer}, {Mocanu}, {Mohr}, {Murray}, {Natoli}, {Padin},
  {Plagge}, {Pryke}, {Rest}, {Ruel}, {Ruhl}, {Saliwanchik}, {Saro}, {Sayre},
  {Schaffer}, {Shaw}, {Shirokoff}, {Song}, {Spieler}, {Staniszewski}, {Stark},
  {Story}, {Stubbs}, {{\v S}uhada}, {van Engelen}, {Vanderlinde}, {Vieira},
  {Vikhlinin}, {Williamson}, {Zahn}, \& {Zenteno}}]{rei13}
{Reichardt}, C.~L., {Stalder}, B., {Bleem}, L.~E., {et~al.} 2013, \apj, 763,
  127

\bibitem[{{Rephaeli}(1980)}]{rap80}
{Rephaeli}, Y. 1980, \apj, 241, 858

\bibitem[{{Rephaeli}(1995)}]{rep95}
{Rephaeli}, Y. 1995, \araa, 33, 541

\bibitem[{{Roth} \& {Bauer}(1999)}]{rot99}
{Roth}, K.~C. \& {Bauer}, J.~M. 1999, \apjl, 515, L57

\bibitem[{{Roth} {et~al.}(1993){Roth}, {Meyer}, \& {Hawkins}}]{rot93}
{Roth}, K.~C., {Meyer}, D.~M., \& {Hawkins}, I. 1993, \apjl, 413, L67

\bibitem[{{Song} {et~al.}(2012){Song}, {Zenteno}, {Stalder}, {Desai}, {Bleem},
  {Aird}, {Armstrong}, {Ashby}, {Bayliss}, {Bazin}, {Benson}, {Bertin},
  {Brodwin}, {Carlstrom}, {Chang}, {Cho}, {Clocchiatti}, {Crawford}, {Crites},
  {de Haan}, {Dobbs}, {Dudley}, {Foley}, {George}, {Gettings}, {Gladders},
  {Gonzalez}, {Halverson}, {Harrington}, {High}, {Holder}, {Holzapfel},
  {Hoover}, {Hrubes}, {Joy}, {Keisler}, {Knox}, {Lee}, {Leitch}, {Liu},
  {Lueker}, {Luong-Van}, {Marrone}, {McDonald}, {McMahon}, {Mehl}, {Meyer},
  {Mocanu}, {Mohr}, {Montroy}, {Natoli}, {Nurgaliev}, {Padin}, {Plagge},
  {Pryke}, {Reichardt}, {Rest}, {Ruel}, {Ruhl}, {Saliwanchik}, {Saro}, {Sayre},
  {Schaffer}, {Shaw}, {Shirokoff}, {{\v S}uhada}, {Spieler}, {Stanford},
  {Staniszewski}, {Stark}, {Story}, {Stubbs}, {van Engelen}, {Vanderlinde},
  {Vieira}, {Williamson}, \& {Zahn}}]{son12}
{Song}, J., {Zenteno}, A., {Stalder}, B., {et~al.} 2012, \apj, 761, 22

\bibitem[{{Songaila} {et~al.}(1994{\natexlab{a}}){Songaila}, {Cowie}, {Hogan},
  \& {Rugers}}]{son94b}
{Songaila}, A., {Cowie}, L.~L., {Hogan}, C.~J., \& {Rugers}, M.
  1994{\natexlab{a}}, \nat, 368, 599

\bibitem[{{Songaila} {et~al.}(1994{\natexlab{b}}){Songaila}, {Cowie}, {Vogt},
  {Keane}, {Wolfei}, {Hu}, {Oren}, {Tytleri}, \& {Lanzetta}}]{son94a}
{Songaila}, A., {Cowie}, L.~L., {Vogt}, S., {et~al.} 1994{\natexlab{b}}, \nat,
  371, 43

\bibitem[{{Srianand} {et~al.}(2004){Srianand}, {Chand}, {Petitjean}, \&
  {Aracil}}]{sri04}
{Srianand}, R., {Chand}, H., {Petitjean}, P., \& {Aracil}, B. 2004, Physical
  Review Letters, 92, 121302

\bibitem[{{Srianand} {et~al.}(2008){Srianand}, {Noterdaeme}, {Ledoux}, \&
  {Petitjean}}]{sri08}
{Srianand}, R., {Noterdaeme}, P., {Ledoux}, C., \& {Petitjean}, P. 2008, \aap,
  482, L39

\bibitem[{{Srianand} {et~al.}(2000){Srianand}, {Petitjean}, \&
  {Ledoux}}]{sri00}
{Srianand}, R., {Petitjean}, P., \& {Ledoux}, C. 2000, \nat, 408, 931

\bibitem[{{Sunyaev} \& {Zel'dovich}(1972)}]{sun72}
{Sunyaev}, R.~A. \& {Zel'dovich}, Y.~B. 1972, Comments on Astrophysics and
  Space Physics, 4, 173

\bibitem[{{Thaddeus}(1972)}]{tha72}
{Thaddeus}, P. 1972, \araa, 10, 305

\bibitem[{{The Dark Energy Survey Collaboration}(2005)}]{des05}
{The Dark Energy Survey Collaboration}. 2005, ArXiv Astrophysics e-prints

\bibitem[{{Tonry} {et~al.}(2012){Tonry}, {Stubbs}, {Lykke}, {Doherty},
  {Shivvers}, {Burgett}, {Chambers}, {Hodapp}, {Kaiser}, {Kudritzki},
  {Magnier}, {Morgan}, {Price}, \& {Wainscoat}}]{ton12}
{Tonry}, J.~L., {Stubbs}, C.~W., {Lykke}, K.~R., {et~al.} 2012, \apj, 750, 99

\bibitem[{{Uzan}(2011)}]{uza11}
{Uzan}, J.-P. 2011, Living Reviews in Relativity, 14, 2

\bibitem[{{Uzan} {et~al.}(2004){Uzan}, {Aghanim}, \& {Mellier}}]{uza04}
{Uzan}, J.-P., {Aghanim}, N., \& {Mellier}, Y. 2004, \prd, 70, 083533

\bibitem[{{Zel'dovich} \& {Sunyaev}(1969)}]{zel69}
{Zel'dovich}, Y.~B. \& {Sunyaev}, R.~A. 1969, \apss, 4, 301

\end{thebibliography}

\end{document}